\documentclass[english]{article}

\usepackage[colorlinks,bookmarksopen,bookmarksnumbered,allcolors=red]{hyperref}
\usepackage{xcolor}
\usepackage{graphicx}
\usepackage{geometry}
\usepackage{amsmath, amssymb, bm}
\usepackage{authblk}
\usepackage{placeins}
\usepackage{pdflscape}
\usepackage{rotating}
\usepackage{multirow}
\usepackage{booktabs}
\usepackage{float}
\usepackage{caption}
\usepackage{siunitx} 
\usepackage{subcaption}
\usepackage{mathrsfs}
\definecolor{bluegreen}{RGB}{46,141,131}
\definecolor{darkgreen}{RGB}{46,139,87}
\definecolor{darkred}{RGB}{219,7,61}
\definecolor{darkblue}{RGB}{0,0,137}

\title{{\textbf{Bayesian Joint Modelling of Longitudinal Creatinine Trajectories in Children with Auto-Immune Disorders to Predict Paediatric Kidney Disease Risk in a Single Centre Study}}}%Predicting Paediatric Kidney Disease Risk from Longitudinal Creatinine Measurements: A Joint Modelling Study Using Great Ormond Street Hospital Data}}}
\author[1]{Qendresa Selimi}
\author[1]{Christiana Charalambous} 

\author[,1]{Taban Baghfalaki\thanks{Corresponding author: taban.baghfalaki@manchester.ac.uk}}
\author[2]{John Booth}
\author[3,4]{Stephen D Marks}
\affil[1]{Department of Mathematics, The University of Manchester, Manchester, UK}
\affil[2]{Clinical Informatics Research Programme, Great Ormond Street Hospital for Children NHS Foundation Trust, London, UK}
\affil[3]{Department of Paediatric Nephrology, Great Ormond Street Hospital for Children NHS Foundation Trust, London, UK}
\affil[4]{NIHR Great Ormond Street Hospital Biomedical Research Centre, University College London Great Ormond Street Institute of Child Health, London, UK}

\date{}

\begin{document}

\maketitle

\begin{abstract}
This study investigates the relationship between longitudinal serum creatinine measurements and the risk of adverse kidney outcomes in paediatric patients with auto-immune disorders at Great Ormond Street Hospital for Children NHS Foundation Trust, London. To jointly analyse repeated biomarker measurements and time-to-event outcomes, we employed a joint modelling framework that combines the creatinine trajectories with  the time to death or diagnosis of acute kidney injury  or chronic kidney disease. Covariates considered in analysis included demographic and clinical characteristics.
The results demonstrate a strong association between evolving creatinine profiles and the risk of the composite event. Specifically, treatment with corticosteroids and calcium channel blockers was associated with an increased event risk, whereas immunosuppressive therapy was associated with a reduced risk. The longitudinal component showed that creatinine trajectories were significantly influenced by age and BMI z-score.
To demonstrate the practical utility of the proposed framework, dynamic  risk predictions were generated using patients’ observed creatinine trajectories. Model performance was compared using  model selection criteria, alongside area under the curve  and Brier score  to evaluate the accuracy of dynamic risk predictions. These predictions illustrate the potential of joint models to support personalised medicine and clinical decision making in paediatric nephrology through real-time risk assessment. 
\end{abstract}

\section{Introduction}\label{sec1}
Inflammatory diseases are conditions characterised by abnormal activity of the immune system, leading to persistent inflammation that may damage multiple organs, including the kidneys. These diseases generally occur in two main forms: autoimmune diseases, such as Systemic Lupus Erythematosus (SLE), in which the immune system attacks the body's own tissues, and vasculitis, which refers to inflammation of the blood vessels. When the kidneys are affected, serious impairment of kidney function may occur, presenting as either acute kidney injury (AKI) or chronic kidney disease (CKD), depending on the timing of the inflammation. Both AKI and CKD are serious complications that require continuous monitoring to prevent subsequent kidney failure or damage, causing end-stage kidney disease (ESKD), which is the most end of the spectrum of CKD necessitating kidney replacement therapy with dialysis and/or transplantation.\\
Kidney disease is a major global health issue \cite{bibkov_et_al_2019}. It is estimated that more than 800 million people worldwide are affected by CKD \cite{kovesdy_2022,jager_et_al_2019}.  Over one million deaths were attributed to CKD in 2017 alone \cite{bibkov_et_al_2019}. Furthermore, a study of the United States of America population reported approximately 125,000 recorded cases of ESKD in 2017, with an increasing trend in CKD prevalence over time \cite{saran_et_al_2020}.
The management of kidney disease is particularly challenging in children compared with adults. Children with kidney disease often present with congenital anomalies of the kidneys and urinary tract (CAKUT) and different clinical manifestations, which may lead to delayed or inaccurate diagnosis. Research on the impact of auto-immune disorders on kidney function in children remains relatively limited \cite{children_kidney}.\\
Early research on kidney disease progression often relied on traditional survival models such as the Cox proportional hazards model \cite{cox_1972}. Although well suited to time to event outcomes (e.g. death or onset of ESKD), these models do not fully exploit the information contained in longitudinal biomarkers, such as serum creatinine, that evolve over time. Both longitudinal and survival outcomes are collected on the same subjects and so are typically correlated; analysing them separately can lead to biased or inefficient estimates \cite{guo2004separate}. Furthermore, naively treating biomarker measurements as fixed or time-dependent covariates, may still fail to capture the underlying trajectory of kidney function over time. To address this limitation, longitudinal modelling approaches, particularly 
linear mixed effects models (LME) \cite{laird1982random,verbeke2000}, were introduced to analyse repeated biomarker measurements and capture subject-specific trajectories. However, these so called two-stage approaches \cite{albert2010estimating}, where predicted biomarker values as opposed to the raw values were used in the time-to-event model, remained problematic, resulting in underestimation of the association between the longitudinal and time-to-event outcomes. \\
This motivated the development of joint models for longitudinal and time-to-event data \cite{rizopoulos2012joint}, which simultaneously model the biomarker trajectory and the time-to-event process while accounting for both measurement error and their underlying association \cite{wulfsohn1997,tsiatis2004,sayyadi2017assessing}. These approaches have become widely used in studies of renal disease progression, where longitudinal biomarkers such as serum creatinine or estimated glomerular filtration rate (eGFR) are analysed jointly with time-to-event outcomes, including kidney failure, transplantation, or death. Joint models also enable dynamic risk prediction, providing clinically valuable insight into how changes in kidney biomarkers relate to CKD.\cite{liao2024approach,rizopoulos2012joint}.\\
Unlike previous studies, which have primarily focused on adults or analysed longitudinal and survival outcomes separately, our study applies a joint modelling framework specifically to a paediatric cohort of auto-immune disorders in the UK. To the best of our knowledge, there is very little work in the literature focusing on joint models for renal disease in children. 
Most existing work focuses on the chronic kidney disease in children (CKiD) study based in the United States of America and Canada (\cite{atkinson2021ckid, ng2019incidence, weidemann2020plasma}), and uses survival models to analyse kidney disease progression. 
However, 
Armero et al. \cite{armero2018} proposed a Bayesian joint model to assess the progression of CKD in children, linking longitudinal measures of kidney function with survival outcomes such as disease progression events. This work illustrates the feasibility and relevance of joint modelling approaches in paediatric renal research, where repeated measurements and time‑to‑event outcomes are both of interest. \\
In this study, we demonstrate how joint modelling can be used to investigate the relationship between longitudinal serum creatinine trajectories and the risk of AKI or CKD in children with auto-immune disorders.  In addition, by modelling the cumulative exposure to elevated creatinine over time, we provide clinically interpretable risk estimates that can be updated as new measurements become available. This approach addresses an important gap in the paediatric nephrology literature and offers a practical statistical tool for real-time clinical decision-making. \\%In this study, we demonstrate how joint modelling can be used to investigate the relationship between serum creatinine levels and the risk of developing AKI or CKD in children. In addition, we perform dynamic risk prediction based on repeated creatinine measurements. Our approach addresses a gap in the literature and provides a practical tool for real-time clinical decision-making in children. 
The remainder of the paper is organised as follows. Section 2 describes the study cohort including the baseline patient characteristics, the longitudinal serum creatinine measurements, the time-to-event outcome, and the covariates considered in the analysis. Section 3 presents the statistical methodology, beginning with the longitudinal and survival sub-models, followed by the proposed joint modelling framework that links repeated creatinine measurements with the risk of the composite renal event.  Section 4 presents the data analysis, including the specification of the submodels. It also describes the method used to handle the missing time-dependent BMI z-score covariate and outlines the dynamic prediction framework. We report the predictive performance measures used to assess the model, the criteria for model selection, and the posterior parameter estimates. Furthermore, we interpret the association between creatinine trajectories and event risk and illustrate subject-specific dynamic risk predictions. Finally, Section 5 concludes with the main findings, their clinical relevance, the limitations of the study, and directions for future research.

\section{Data Description}\label{sec2}
We used data from Great Ormond Street Hospital NHS Foundation Trust (GOSH) in London, a leading paediatric nephrology centre, comprising 514 paediatric patients of the GOSH Paediatric Kidney Cohort (GPKC) with a total of $8702$ repeated measurements. The dataset included information on demographic characteristics, diagnoses, laboratory test results, and medications. Patients were followed from 1 April 2019 to 31 December 2025, entering the study at varying times during the follow-up period.

\begin{figure}[t!]
    \centering
    \includegraphics[width=0.6\linewidth]{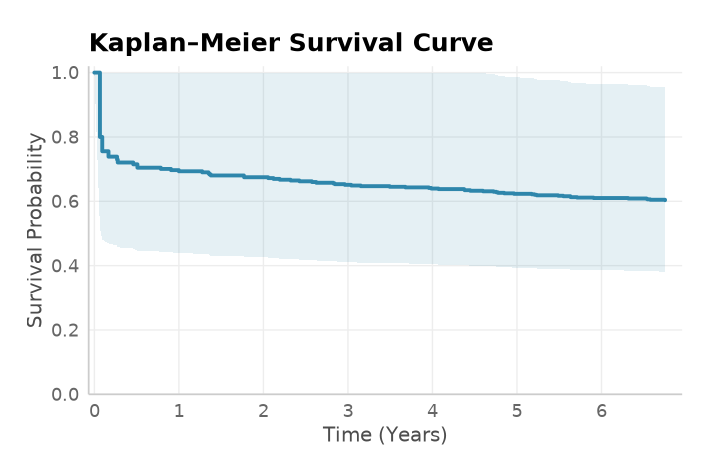}
\caption{Kaplan–Meier survival curve estimating the probability of remaining free from acute kidney injury (AKI), chronic kidney disease (CKD), or death over time in the GPKC dataset.}\label{km}
\end{figure}

\paragraph{Baseline Characteristics and Covariates}
We considered variables that may influence creatinine levels and the risk of kidney disease, including age at study entry, sex, body composition, comorbid conditions, renal history, and treatment regimes. The mean age at study entry was 11.1 years (SD = 3.96), with a range from one week to 17.96 years. Most patients were between 10 and 15 years old (52.7\%), followed by 5–10 years (22.4\%), 15–18 years (15.4\%), and 0–5 years (9.5\%). To improve model convergence and numerical stability, age at entry was standardised.\\
Body Mass Index (BMI) is a measure of body composition accounting for weight relative to height. In our cohort, the mean (SD) BMI was 21 (5.4), ranging from 10.5 to 44. As BMI is more variable in children, we used BMI z-scores, which adjust BMI for age and sex relative to a reference population. BMI z-scores are commonly calculated using the UK 1990 (UK90) growth reference in the UK \cite{cole1998british}. 
BMI measurements had missing values at some timepoints. To address this issue, Section \ref{model} outlines the approach considered for handling missing time-dependent covariates.\\
%To address this, imputed BMI values at the creatinine measurement timepoints using a linear mixed effects model. %(formula: $BMI_{ij} =  \beta_{0} + \beta_{1}\mathrm{age}_{i}+ \beta_{2}\mathrm{sex}_{i}+\beta_3\mathrm{s}_{ij}+b_{0i}+b_{1i}\mathrm{s}_{ij}+\epsilon_{ij}$). Subsequently, BMI z-scores were calculated from the imputed BMI values using the R package \texttt{childsds} \cite{cole1998british}.
We also included two clinically important binary covariates in the analysis. The first indicates the presence of comorbid conditions, specifically hypertension (53 patients), diabetes mellitus (9 patients), or hepatitis (3 patients). Patients were coded as having a comorbidity if at least one of these conditions was present, and as not having a comorbidity otherwise. The second covariate captures the presence of pre-existing kidney-related abnormalities in patients without a prior AKI or CKD diagnosis. These included conditions such as renal cysts, nephrolithiasis, renal hypoplasia or dysplasia, renal artery stenosis, and other similar renal abnormalities. Patients were coded as positive for kidney-related conditions if any of these abnormalities were present.  
%We also define two key covariates in our study: \texttt{Comorbidity} and \texttt{Kidney\_condition}. \texttt{Comorbidity} indicates the presence of any of the following conditions: diabetes (9 patients), hypertension (53 patients), or hepatitis (3 patients). A patient is coded as \texttt{Comorbidity} = 1 if at least one of these conditions is present, and 0 otherwise. \texttt{Kidney\_condition} captures the presence of other kidney-related issues, even in patients without an AKI/CKD diagnosis. These conditions include kidney cysts, kidney calculus, renal hypoplasia, congenital renal artery stenosis, and similar abnormalities. A patient is coded as \texttt{Kidney\_condition} = 1 if any of these conditions are present, and 0 otherwise.
In addition, the study included a total of 11 commonly used medications, classified into two main therapeutic categories, namely autoimmune and cardiovascular, and further divided into six subgroups based on their mechanism of action. For modelling purposes, a binary covariate was created for each subgroup, indicating whether a patient received any medication within that subgroup. Table~\ref{medications} in Appendix A provides a summary of the medication groups, subgroups, and specific drugs. \\ % A complete summary of baseline characteristics and all binary covariates is presented in Table~\ref{covariates}.
Table~\ref{covariates} summarises baseline characteristics and treatment exposures of the study population. The majority of patients were female (61\%), and relatively few had recorded comorbidities (13\%) or pre-existing kidney conditions (9\%). Corticosteroids were the most commonly prescribed treatment (58\%), whereas cardiovascular medications (including anti-hypertensive medications) were less frequent (17–18\%). 
\begin{table}[ht]
\centering
\footnotesize
%\caption{Baseline characteristics and covariates of the study population ($N = 514$). Percentages are shown in parentheses.}
\caption{Baseline characteristics and key covariates of the GPKC dataset. Percentages are shown in parentheses.}\label{covariates}
\begin{tabular}{lll}
\toprule
\textbf{Variable} & \textbf{Category} & \textbf{Count (\%)} \\
\midrule

\textbf{Sex} & Male   & 201 (39\%) \\
             & Female & 313 (61\%) \\[2pt]

\textbf{Comorbidity} & Yes & 67 (13\%) \\
                     & No  & 447 (87\%) \\[2pt]

\textbf{Kidney Condition} & Yes & 47 (9\%) \\
                          & No  & 467 (91\%) \\[2pt]

\textbf{Autoimmune Medications} & & \\
Corticosteroid  & Yes & 299 (58\%) \\
                  & No  & 215 (42\%) \\
Immunosuppressant  & Yes & 189 (37\%) \\
                      & No  & 325 (63\%) \\
Immune modulator  & Yes & 139 (27\%) \\
                     & No  & 375 (73\%) \\
B-cell depletion therapy & Yes & 120 (23\%) \\
               & No  & 394 (77\%) \\[2pt]

\textbf{Hypertension / Cardiovascular Medications} & & \\
Calcium channel blocker  & Yes & 87 (17\%) \\
                        & No  & 427 (83\%) \\
ACE inhibitor  & Yes & 95 (18\%) \\
             & No  & 419 (82\%) \\
\bottomrule
\end{tabular}
\end{table}
\paragraph{Time-to-Event Outcome} We defined a composite event consisting of death, diagnosis with AKI, or diagnosis with CKD. The event time is the earliest occurrence of any of these outcomes, measured in years from the study origin (1 April 2019) to the date of diagnosis, with subjects entering the risk set at the time of their first serum creatinine measurement. Patients who did not experience any event were censored at the end of follow-up (31 December 2025). Out of 514 patients, 57 (11\%) experienced an event: 36 with AKI, 15 with CKD, and 6 deaths. \\
Figure~\ref{km} presents the Kaplan–Meier curve, estimating the probability of remaining free from the composite event over time. As we can see, after an initial 20\% drop within the first month of followup, the survival probability of the cohort slowly decreases and plateaus around 60\%. Understanding the relationship between creatinine trajectories and event risk is clinically important, as it may help prevent end-stage kidney disease and guide individualised treatment strategies. 

\FloatBarrier

\begin{figure}
    \centering
    \includegraphics[width=0.6\linewidth]{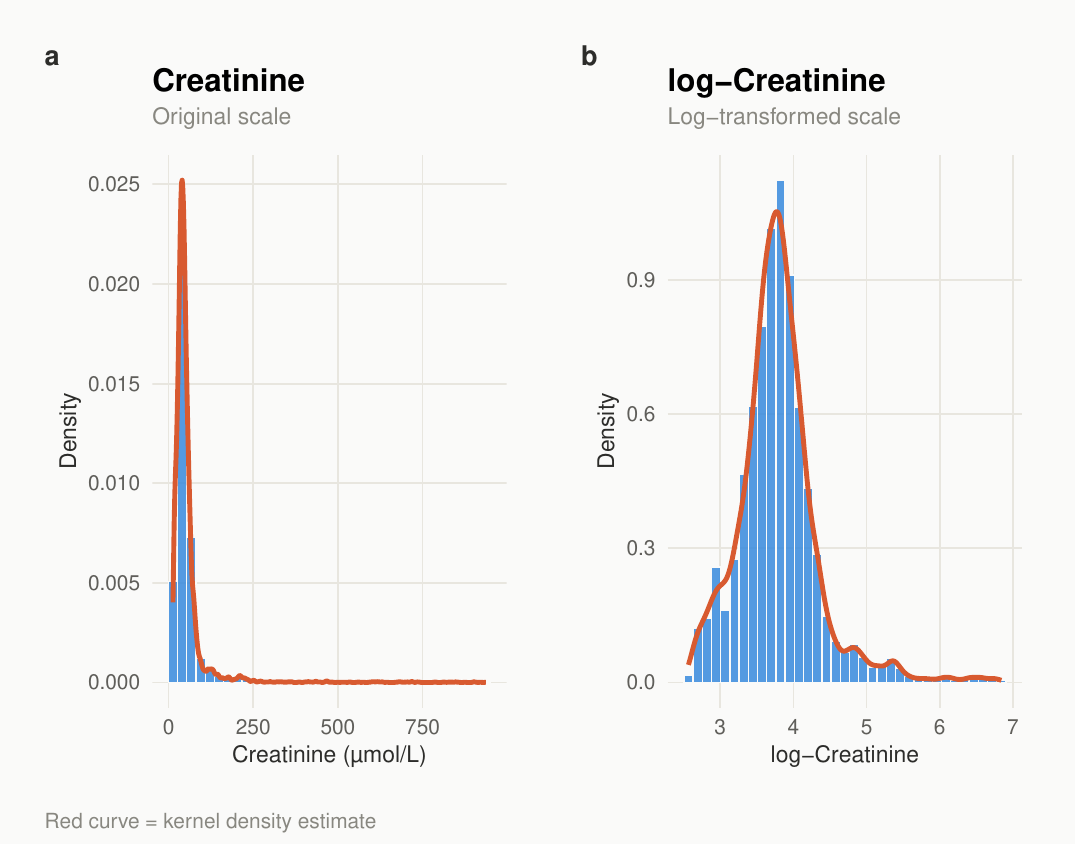}
\caption{Distribution of serum creatinine and its log-transformed values in the GPKC dataset.}\label{hist}
\end{figure}

\begin{figure}[t!]
    \centering
    \includegraphics[width=16cm]{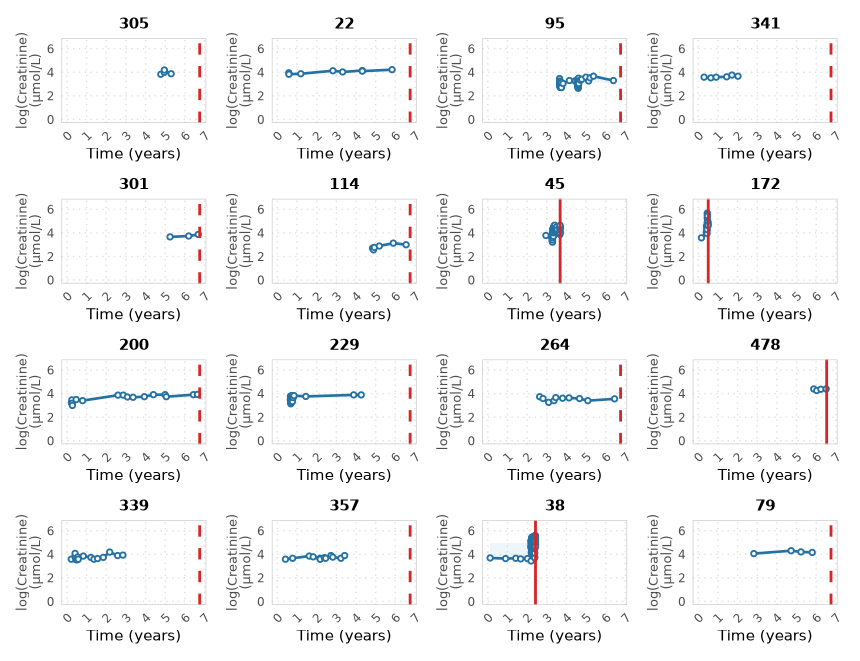}
    \caption{Log-transformed serum creatinine trajectories for a randomly selected subset of patients from  the GPKC dataset. Solid red lines represent patients who experienced the composite event, while dashed red lines represent censored patients.}
    \label{panel}
\end{figure}

\paragraph{Longitudinal Creatinine Measurements}
Serum creatinine levels were obtained from routine laboratory tests. Patients had a mean of 17 longitudinal measurements each, ranging from 2 to 313; those with only a single measurement were excluded from the analysis. Panel (a) of Figure \ref{hist} presents a histogram of the raw creatinine values, which showed substantial variability across the cohort, with a mean (SD) of 54.5 (64.8). To improve numerical stability and account for skewness in the distribution, creatinine values were log-transformed. Panel (b) of Figure \ref{hist} displays the histogram of the log-transformed creatinine values, demonstrating a more symmetric distribution suitable for modelling.\\
Figure~\ref{panel} displays the individual log-transformed creatinine trajectories for a randomly selected subset of patients. Solid red lines indicate patients who experienced the composite event (AKI, CKD, or death), while dashed red lines indicate censored patients. The figure highlights substantial variability in both the number and timing of longitudinal measurements across patients. For instance, patients 45, 38, and 172 had a large number of repeated measurements near the critical event time, with many showing noticeable increases in creatinine levels. Conversely, for several patients (e.g., 341, 229, and 339), there were substantial gaps between the last measurement and the event or censoring time, illustrating irregular follow-up patterns.

\FloatBarrier

\FloatBarrier

\section{Statistical Methods for Joint Analysis}\label{Methodology}
In this study, we investigate the association between longitudinal serum creatinine measurements and the risk of a composite clinical event (AKI, CKD, or death) in paediatric patients. To achieve this, we employ a joint modelling framework that simultaneously accounts for repeated measurements of creatinine and the time-to-event outcome. This approach allows proper handling of within-patient correlation, variability in measurement frequency and timing, and the quantification of how changes in creatinine trajectories influence the hazard of experiencing the clinical event \cite{rizopoulos2012joint}.\\
The joint modelling framework consists of two interconnected sub-models: a longitudinal sub-model for repeated creatinine measurements and a survival sub-model for the time-to-event outcome. These sub-models are linked through the longitudinal biomarker, with the survival model depending on the current value, slope, cumulative history, or a linear combination of these creatinine measures for each patient, in addition to baseline covariates. In the following subsections, we describe each component in detail.
\subsection*{Longitudinal submodel}
Longitudinal data are often collected repeatedly over time for each subject. Let $y_{ij}=y_i(s_{ij})$ denote the $j$-th measurement at time $s_{ij}$ for subject $i$, where $i=1,\ldots,n$ and $j=1,\ldots,n_i$. A widely used approach for analysing such repeated measurements is the linear mixed effects model \cite{laird1982random}, defined as
\begin{equation}\label{long_model}
y_i(s_{ij}) \mid (\bm{X}_{li}, \bm{Z}_i, \bm{b}_i) = 
\eta_i(s_{ij}) + \epsilon_{ij} 
= \bm{X}_{li}(s_{ij})^\top \bm{\beta} + \bm{Z}_i^\top(s_{ij}) \bm{b}_i + \epsilon_{ij}, 
\quad \epsilon_{ij} \sim N(0, \sigma^2),
\end{equation}
where $\bm{X}_{li}(s_{ij})$ is the vector of fixed-effect covariates with associated coefficients $\bm{\beta}$, $\bm{Z}_i(s_{ij})$ is the vector of covariates associated with the subject-specific random effects $\bm{b}_i \sim N(\bm{0}, \bm{D})$, and $\epsilon_{ij}$ represents independent measurement errors. The linear predictor $\eta_i(s_{ij})$ represents the underlying trajectory of the longitudinal outcome for subject $i$ at time $s_{ij}$.

\subsection*{Survival submodel}
%Time to event data measure the time to an event of interest. In a medical context, the outcome of interest is usually a clinical event, hence time to event data are commonly referred to as survival data. 
%Let $T_i^{*}$ denote the event time for subject $i$ and $C_{i}$ be the censoring time.  We only observe $T_{i}=\textup{min}(T^{*}_{i}, C_{i})$, and let $\delta _{i}=\mathbb{I}(T_{i}^{*}\leq C_{i})$ be the event indicator for subject $i$.
%Survival data commonly models the hazard of experiencing the event via a Cox Proportional hazards model, \cite{cox_1972}, as follows: 
%\begin{equation*}
%\lambda _{i}(t|, X_i, Z_i, X_{2i}, \bm{b}_i)=\lambda _{0}(t)\exp(\bm{X}_{si}(t)^\top \bm{\omega}+\alpha \cdot \eta_i(t))
%\end{equation*}
% where $\lambda_i(t)$ is the hazard of subject $i$ experiencing the event at time $t$, $\lambda_{0}(t)$ is the baseline hazard at time $t$, and $X_{si}$ is the covariate matrix with corresponding coefficients $\omega$. Additionally, we assume $\delta_i$ is the event indicator for subject $i$, such that $\delta_i=1$ if the event occurred and 0 otherwise.
Let $T_i^{*}$ denote the true event time for subject $i$, and let $C_i$ denote the censoring time. The observed follow-up time is defined as $T_i=\min(T_i^{*},C_i)$, while $\delta_i=\mathbb{I}(T_i^{*}\leq C_i)$, where $\delta_i=1$ if the event is observed and $\delta_i=0$ otherwise.\\
Time-to-event outcomes are commonly modelled using a Cox proportional hazards model \cite{cox_1972}. Within the joint modelling framework, the hazard function for subject $i$ at time $t$ is specified as
\begin{equation}\label{surv_model}
\lambda_i(t \mid \bm{X}_{si}, \mathcal{M}_i(t)) =\lambda_0(t)\exp\left\{
\bm{X}_{si}^{\top}\bm{\omega}+ g\left(\mathcal{M}_i(t),\bm{\alpha}\right)
\right\},
\end{equation}
where $\lambda_i(t)$ denotes the instantaneous risk of experiencing the event at time $t$, $\lambda_0(t)$ is the baseline hazard function, and $\bm{X}_{si}$ is the vector of baseline covariates with corresponding regression coefficients $\bm{\omega}$.\\
The term $g(\mathcal{M}_i(t),\bm{\alpha})$ defines the association structure linking the longitudinal and survival processes through the subject-specific longitudinal trajectory $\mathcal{M}_i(t)$. Common choices include the current value $\eta_i(t)$, the current slope $\eta_i'(t)$, or cumulative exposure $\int_0^t \eta_i(u)\,du$, with corresponding association parameters collected in $\bm{\alpha}$. More generally, the association structure may be specified as a linear combination of these components, for example
$g(\mathcal{M}_i(t),\bm{\alpha})=\alpha_1 \eta_i(t) +\alpha_2 \eta_i'(t)$,
allowing the event hazard to depend simultaneously on both the biomarker level and its rate of change over time.
This association structure assumes that both the current underlying biomarker value and its instantaneous rate of change at the same time point contribute to the hazard of the event. 
The parameter $\alpha_2$ can be interpreted as the change in the log-hazard of the event at time $t$ associated with a one-unit increase in the slope of the biomarker trajectory at time $t$, while holding the current biomarker value $\eta_i(t)$ constant.  
In contrast, the cumulative exposure association structure assumes that the hazard of the event at time $t$ depends on the total accumulated exposure of the longitudinal biomarker up to that time.

\subsection{Parameter estimation}
Estimation of joint models can be performed within both frequentist and Bayesian frameworks. In the frequentist framework, parameters are typically estimated via maximum likelihood using the EM algorithm \cite{rizopoulos2012joint}, which integrates over the random effects to jointly model the longitudinal and survival outcomes.\\
Let $\bm{y}_i = (y_{i1}, \ldots, y_{in_i})^\top$ denote the longitudinal measurements for subject $i$, and let $\bm{\theta}$ denote the collection of all model parameters. Define the observed data as
$\mathcal{D} = \{T_i, \delta_i, y_{ij} : j = 1, \ldots, n_i, \; i = 1, \ldots, n\}$.
Then, the likelihood function for the joint model, combining the longitudinal sub-model \eqref{long_model} and the survival sub-model \eqref{surv_model}, is
\begin{align}\label{joint_likelihood}
\mathcal{L}(\bm{\theta} \mid \mathcal{D}) &= 
\prod_{i=1}^{n} \int 
f(\bm{y}_i, T_i, \delta_i \mid \bm{b}_i, \bm{\theta}) \, 
f(\bm{b}_i \mid \bm{\theta}) \, d\bm{b}_i  \\
&\quad =
\prod_{i=1}^{n} \int 
f(\bm{y}_i \mid \bm{b}_i, \bm{\theta}_l) \, 
f(T_i, \delta_i \mid \bm{b}_i, \bm{\theta}_s) \, 
\phi(\bm{b}_i; \bm{0}, \bm{D}) \, d\bm{b}_i \nonumber
\end{align}
where $\bm{b}_i$ are the subject-specific random effects, 
$\bm{\theta}_l$ denotes the parameters of the longitudinal sub-model and 
$\bm{\theta}_s$ denotes the parameters of the survival sub-model. 
Here, $f(\bm{y}_i \mid \bm{b}_i, \bm{\theta}_l)$ is the longitudinal likelihood, 
$f(T_i, \delta_i \mid \bm{b}_i, \bm{\theta}_s)$ is the survival likelihood contribution, 
and $\phi(\bm{b}_i; \bm{0}, \bm{D})$ is the density of the random effects, 
assumed to follow a multivariate normal distribution with mean $\bm{0}$ and covariance matrix $\bm{D}$.\\
By assuming conditional independence of the longitudinal and survival outcomes given the random effects, Equation \eqref{joint_likelihood} can be rewritten as
\begin{equation}\label{joint_likelihood_cond}
\mathcal{L}(\bm{\theta} \mid \mathcal{D}) = \prod_{i=1}^{n} \int 
\Big[\Big(\prod_{j=1}^{n_i} f(y_{ij} \mid \bm{b}_i, \bm{\theta}_l)\Big) \, f(T_i, \delta_i \mid \bm{b}_i, \bm{\theta}_s) \, \phi(\bm{b}_i; \bm{0}, \bm{D})\Big] \, d\bm{b}_i.
\end{equation}
This formulation explicitly highlights that, conditional on the random effects $\bm{b}_i$, the longitudinal measurements and the survival outcome are independent.\\
Because this integral often has no closed form, numerical methods such as Gaussian quadrature, Laplace approximation, or the EM algorithm are used \cite{rizopoulos2012joint, tsiatis2004joint}. This approach provides maximum likelihood estimates of the fixed effects, variance components, and association parameters, with well-established asymptotic properties.\\
In a Bayesian context, the subject-specific random effects $\bm{b}_i$ are also treated as parameters. The joint posterior distribution of the parameters and random effects can then be expressed as:
\begin{equation*}
\pi(\bm{\theta}, \{\bm{b}_i\}_{i=1}^n \mid \mathcal{D}) \propto \mathcal{L}(\bm{\theta} \mid \mathcal{D}) \, \phi(\bm{b}_i; \bm{0}, \bm{D})  \, \pi(\bm{\theta}),
\end{equation*}
where   $\bm{\theta} = \{\bm{\beta}, \bm{\omega}, \alpha, \sigma^2, \mathbf{D}\}$ represents all model parameters and $\pi(\bm{\theta})$ denotes the joint prior distribution of all model parameters \cite{papageorgiou_et_al_2019}.   For the priors, we specify:
\begin{align*}
\bm{\beta} &\sim N(\bm{0}, \sigma_\beta^2 \mathbf{I}),   \\
\bm{\omega} &\sim N(\bm{0}, \sigma_\omega^2 \mathbf{I}), \\
\bm{\alpha} &\sim N(0, \sigma_\alpha^2\mathbf{I}),  \\
\sigma^2 &\sim \text{Inverse-Gamma}(a,b),  \\
\mathbf{D} &\sim \text{Inverse-Wishart}(\nu, \mathbf{S}),  
\end{align*}
Here, $N(\cdot,\cdot)$ denotes a multivariate normal distribution, with dimension corresponding to the number of parameters in $\bm{\beta}$, $\bm{\omega}$, or $\bm{\alpha}$, and a diagonal covariance matrix. $\text{Inverse-Gamma}(a,b)$ and $\text{Inverse-Wishart}(\nu, \mathbf{S})$ denote the standard inverse-gamma and inverse-Wishart distributions, respectively.
\\
We fit the joint model to our renal cohort using the \verb|JMBayes2| R package \cite{jmbayes2}, which implements a Markov Chain Monte Carlo (MCMC) algorithm to approximate the posterior distribution. Posterior summaries, including means, standard deviations, and credible intervals, are then obtained for both fixed and random effect parameters, allowing probabilistic inference on the association between longitudinal creatinine trajectories and the risk of the composite event.

\subsection{Dynamic risk prediction}
Dynamic risk prediction plays a crucial role in personalised medicine, as it allows clinicians to update a patient’s risk of experiencing an event based on their individual longitudinal biomarker history. Unlike traditional survival analysis, which only considers baseline covariates, dynamic prediction incorporates repeated measurements over time, capturing the evolving trajectory of a patient’s health status. This is particularly important in chronic diseases, such as kidney disease, where biomarker trajectories (e.g., creatinine levels) can provide early warning signals for adverse outcomes \cite{rizopoulos2011dynamic,rizopoulos2017dynamic}. By leveraging longitudinal data, joint models facilitate individualised prognosis, enabling timely interventions and improved clinical decision-making \cite{taylor2013real,rizopoulos2014combining}.\\
In joint modelling, an important quantity of interest is the subject-specific dynamic probability of experiencing the event within a future prediction window, conditional on survival and the observed longitudinal history up to a landmark time $t_L$. Let
$\mathcal{Y}_i(t_L)=\{ y_i(s_{ij}) : 0 \le s_{ij}\le t_L,\; j=1,\ldots,n_i \}$
denote the longitudinal history of subject $i$ up to time $t_L$. For a prediction horizon $\Delta t$, the dynamic event probability is defined as
\begin{equation}
\begin{split}
\pi_i(t_L+\Delta t \mid t_L)
&= P(T_i \le t_L+\Delta t \mid T_i>t_L,\mathcal{Y}_i(t_L),\mathcal{D}) \\
&= 1 -
\iint
\frac{S(t_L+\Delta t \mid \bm{b}_i,\bm{\theta})}
{S(t_L \mid \bm{b}_i,\bm{\theta})}
\, p(\bm{b}_i \mid T_i>t_L,\mathcal{Y}_i(t_L),\bm{\theta})
\, p(\bm{\theta} \mid \mathcal{D})
\, d\bm{b}_i\, d\bm{\theta},
\end{split}
\end{equation}
where $S(\cdot \mid \bm{b}_i,\bm{\theta})$ denotes the subject-specific survival function, $\bm{b}_i$ represents the random effects for subject $i$, and $\bm{\theta}$ denotes the full vector of model parameters.\\
To assess the predictive accuracy of joint models, two commonly used tools are the Area Under the Receiver Operating Characteristic Curve (AUC) and the Brier Score (BS). The AUC measures a model's ability to discriminate between patients who experience the event within a given time frame and those who do not. For a pair of subjects $i$ and $j$, each with longitudinal measurements up to time $t_L$, the AUC can be defined as
\begin{equation}
\text{AUC}(t_L, \Delta t) = \text{P}\Big(
\pi_i(t_L+\Delta t \mid t_L) > \pi_j(t_L+\Delta t \mid t_L)
\;\Big|\;
\{T_i \in (t_L, t_L+\Delta t)\} \cap \{T_j > t_L+\Delta t\}
\Big),
\end{equation}
that is, if subject $i$ experiences the event during the time window while subject $j$ survives, the model should assign a higher probability of experiencing the event to subject $i$ \cite{rizopoulos2017dynamic}.\\
The BS is a measure of prediction error, defined as the mean squared difference between the observed survival status and the predicted survival probability at time $t_L$:
\begin{equation*}
\text{BS}(t_L, \Delta t) = \mathbb{E}\left[\left(\mathbb{I}(t_L<T \leq t_L + \Delta t) - \pi(t_L, \Delta t)\right)^2 \mid T > t_L\right].
\end{equation*}
Due to censoring, direct computation of the indicator function $\mathbb{I}(t_L < T \le t_L + \Delta t)$ is not always possible; therefore, Inverse Probability of Censoring Weighting (IPCW) estimators have been proposed \cite{blanche_et_al_2015}. Based on this approach, we have 
\begin{equation*}
\widehat{\textup{AUC}}(t_L, \Delta t)=\frac{\sum_{i=1}^{n}\sum_{j=1}^{n}\mathbb{I}(\pi_i(t_L, \Delta t)>\pi_j(t_L,  \Delta t))\widetilde{\mathbb{I}}_i(t_L, \Delta t)(1-\widetilde{\mathbb{I}}_j(t_L, \Delta t)) \widehat{W}_i(t_L, \Delta t)\widehat{W}_j(t_L, \Delta t) }{\sum_{i=1}^{n}\sum_{j=1}^{n}\widetilde{\mathbb{I}}_i(t_L, \Delta t)(1-\widetilde{\mathbb{I}}_j(t_L, \Delta t)) \widehat{W}_i(t_L, \Delta t)\widehat{W}_j(t_L, \Delta t)},
\end{equation*}
and 
\begin{equation*}
\widehat{\text{BS}}(t_L, \Delta t) = 
\frac{1}{n\,\widehat{S}_{\widetilde{T}}(t_L)}
\sum_{i=1}^{n}
\widehat{W}_i\,
\left(\tilde{\mathbb{I}}_i - \hat{\pi}_i\right)^{2},
\end{equation*}
where 
\begin{equation*}
\widehat{S}_{\widetilde{T}}(t_L)=\frac{1}{n}\sum_{i=1}^{n}\mathbb{I}(\widetilde{T}_i>t_L), 
\end{equation*}
estimates the probability of surviving up to time $t_L$, and 
\begin{equation*}
\widehat{W}_i(t_L, \Delta t)=\frac{\mathbb{I}(\widetilde{T}_i>t_L+\Delta t)}{\widehat{K}(t_L+\Delta t| t_L)}+
 \frac{\mathbb{I}(t_L<\widetilde{T}_i<t_L+\Delta t)^{\Delta_i}}{\widehat{K}(\widetilde{T}_i| t_L)},
\end{equation*}
with $\widehat{K}(u)$ being the Kaplan-Meier estimator of survival function of the censoring time at $u$.

\section{Joint Model Analysis of the GPKC Dataset}\label{model}
%In this section, we analyse the study data using a joint modelling framework, which simultaneously models the longitudinal trajectory of creatinine measurements and the time-to-event outcome (composite of death, AKI, or CKD). The joint model allows us to investigate the association between changes in creatinine over time and the risk of developing kidney-related events, while appropriately accounting for measurement error in the longitudinal biomarker and censoring in the survival data.
In this section, we analyse the GPKC dataset using a joint modelling framework that simultaneously characterises the longitudinal trajectory of serum creatinine measurements and the time-to-event outcome, defined as the composite of death, AKI, or CKD. This approach enables us to quantify the association between the evolving biomarker profile and the risk of kidney-related events, while appropriately accounting for within-subject correlation, measurement error in the repeated creatinine measurements, and right censoring in the survival outcome. For this purpose,  we first discuss the longitudinal sub-model to investigate the trajectory of serum creatinine over time. In the longitudinal component, BMI z-scores are considered as a time-varying predictor of serum creatinine measurements. However, because BMI z-scores contain missing values at several follow-up times, we propose a linear mixed effects model  for handling missingness in this time-dependent covariate. We then proceed to the specification of the survival sub-model, followed by the estimation of the full joint model and its use for dynamic risk prediction.
In addition, we evaluate the predictive performance of the fitted model through dynamic risk prediction, providing subject-specific estimates of future event risk based on each patient’s observed creatinine history.

\subsection{Longitudinal SubModel}
To describe how creatinine trajectories evolve over time and evaluate the effects of baseline and time-varying characteristics on creatinine levels, we used a linear mixed effects model for log-transformed creatinine. This model allows for subject-specific random effects and provides predictions of creatinine levels at future time points.
The longitudinal sub-model is specified as:
\begin{equation}
\begin{split}
  \log(\text{Creatinine}_i(s_{ij})) &= \eta_i(\text{s}_{ij}) + \epsilon_{ij} \\
  &= \beta_0 +  \beta_1 \,\text{s}_{ij} +\beta_2 \,\text{Sex}_i + \beta_3 \,\text{SAge}_i + \beta_4 \,\text{BMIZ}_i(s_{ij}) +  b_{0i} + b_{1i} \,\text{s}_{ij} + \epsilon_{ij},
\end{split}
\end{equation}
where $\log(\text{Creatinine}_{ij})$ denotes the log-transformed serum creatinine measurement for subject $i$ at time $s_{ij}$, $\text{Sex}_i$ is a binary indicator taking the value 1 for female and 0 for male,  $\text{SAge}_i$ denotes the standardised age at study entry, and $\text{BMIZ}_{ij}$ is the BMI z-score for subject $i$ measured at time $s_{ij}$. The parameters $\beta_0, \ldots, \beta_4$ represent the fixed effects of the model. The terms $b_{0i}$ and $b_{1i}$ are subject-specific random effects corresponding to the random intercept and random slope for time, respectively, assumed to follow a bivariate normal distribution, $(b_{0i}, b_{1i})^\top \sim \mathcal{N}(\bm{0}, \bm{D})$,
where $\bm{D}$ is the covariance matrix that captures between-subject variability in both baseline creatinine levels and their rate of change over time. The residual error term $\epsilon_{ij}$ is assumed to be independently normally distributed as $\epsilon_{ij} \sim \mathcal{N}(0, \sigma^2)$, and independent of the random effects, representing measurement error and unexplained within-subject variability. Finally, $\eta_i(s_{ij})$ denotes the subject-specific linear predictor at time $s_{ij}$.

%where $\log(\text{Creatinine}_{ij})$ is the log-transformed creatinine for subject $i$ at time $\text{s}_{ij}$,  
%$\text{Sex}_i = 1$ for female and 0 for male, $\text{SAge}$ is standardized age, $\text{BMIZ}_{ij}$ is the BMI z-score for subject $i$ at time $\text{s}_{ij}$,  
%$\beta_0, \dots, \beta_4$ are fixed effects, %with $\beta_0$ representing the baseline for the reference group (male, age 0, BMIZ = 0),  
%$b_{0i}$ and $b_{1i}$ are subject-specific random effects, assumed to follow a multivariate normal distribution  
%$(b_{0i}, b_{1i})^\top \sim \mathcal{N}(\bm{0}, \bm{D})$,
%where $\bm{D}$ is the covariance matrix capturing between-subject variability in baseline creatinine and rate of change, 
%$\epsilon_{ij}$ is the residual error, assumed independent and normally distributed  
%$\epsilon_{ij} \sim \mathcal{N}(0, \sigma^2)$, 
%and independent of the random effects, representing measurement error or unexplained variability, and  
%$\eta_i(\text{s}_{ij})$ denotes the linear predictor for subject $i$ at time $\text{s}_{ij}$.

\paragraph{Handling missing values in BMI z-score:}

BMI measurements were not available at all serum creatinine follow-up time points. Since BMI z-score was included as a time-varying covariate in the longitudinal sub-model, imputation was required to align BMI values with the creatinine measurement times. We assume that the missing BMI measurements are missing at random (MAR), meaning that, conditional on the observed data, the probability of missingness does not depend on the unobserved BMI values \cite{little2019statistical}. \\
Let $\{l_{ij}: i = 1,\ldots,n;\ j = 1,\ldots,n_i\}$ denote the observed BMI measurement times, and
$\{s_{ij}: i = 1,\ldots,n;\ j = 1,\ldots,n_i\}$
the corresponding creatinine measurement times. Our objective was to obtain predicted BMI values at the creatinine follow-up times $s_{ij}, ~i = 1,\ldots,n;\ j = 1,\ldots,n_i$.
\\
To achieve this,
BMI scores were modelled using a linear mixed-effects model:
\begin{equation*}
%\text{BMI}_{i}(s_{ij}) = \phi_0 + \phi_1 s_{ij} + \phi_2 \text{SAge}_i + \phi_3 \text{Sex}_i+ u_{0i} + u_{1i} s_{ij} + \varepsilon_{ij},
\text{BMI}_i(l_{ij}) = \phi_0 + \phi_1 l_{ij} + \phi_2 \text{SAge}_i + \phi_3 \text{Sex}_i
+ u_{0i} + u_{1i} l_{ij} + \varepsilon_{ij},
\end{equation*}
where $\text{SAge}_i$ denotes the standardised age at study entry, $\text{Sex}_i$ denotes sex (1 = female, 0 = male), $\phi_0, \dots, \phi_3$ are fixed-effect coefficients, $(u_{0i}, u_{1i})^\top$ are subject-specific random effects assumed to follow a multivariate normal distribution, and $\varepsilon_{ij} \sim \mathcal{N}(0, \sigma^2)$ represents the residual error.\\
The fitted model was then used to predict BMI at the creatinine measurement times $s_{ij}$ for each subject:
\begin{equation*}
\widehat{\text{BMI}}_i(s_{ij}) = \hat \phi_0 + \hat \phi_1 s_{ij} + \hat \phi_2 \text{SAge}_i + \hat \phi_3 \text{Sex}_i
+ \hat u_{0i} + \hat u_{1i} s_{ij}.
\end{equation*}
Finally, predicted BMI z-scores were obtained using the UK90 growth reference \cite{cole1998british}, i.e., $\widehat{\text{BMIZ}}_{ij} = f(\widehat{\text{BMI}}_{ij}, \text{age}_i, \text{Sex}_i)$, where $f(\cdot)$ denotes the UK90 z-score transformation, calculated using the \verb|childsds| package in \texttt{R} \cite{childsds}. 
\\
%Other methods that could be considered for imputation of BMI scores are nearest neighbour imputation, where a missing value is imputed using the nearest available BMI measurement in time, and BMI prediction via height and weight. Since repeated height and weight measurements are available, both of them can be predicted using linear mixed effects models, and BMI computed at the creatinine measurement times $s_{ij}$ using the predicted height and weight values:  $\widehat{\text{BMI}}_i(s_{ij})= \frac{\widehat{\text{Weight}}_i(s_{ij})}{\widehat{\text{Height}}_i(s_{ij})^2}$.\\
In this analysis, we assume that the missing BMI measurements are MAR, as stated previously. For alternative approaches, see \cite{rubin1987multiple}. If this assumption does not hold, more robust inference can be obtained by employing methods that explicitly model the missingness mechanism, such as selection models, pattern mixture models, or joint models that incorporate the missing data process \cite{daniels2008missing}.

\subsection{Survival Submodel}
The hazard of experiencing the composite kidney-related event (death, AKI, or CKD) was modelled using a Cox proportional hazards framework, incorporating both baseline covariates and features derived from the longitudinal creatinine trajectory:
\begin{align}\label{survival_model}
\lambda_i(t) = \lambda_0(t) \exp\Big(&
\omega_1 \text{Comorb}_i
+ \omega_2 \text{KidneyHist}_i
+ \omega_3 \text{Cortico}_i
+ \omega_4 \text{Immuno}_i
+ \omega_5 \text{ImmMod}_i \nonumber \\
&+ \omega_6 \text{BCell}_i
+ \omega_7 \text{CCB}_i 
+ \omega_8 \text{ACEi}_i
+ \alpha_1 \, \eta_i(t) 
+ \alpha_2 \,  \eta^\prime_i(t)  
+ \alpha_3 \, \int_{0}^{t} \eta_i(s)\, ds
\Big),
\end{align}
where \(\lambda_0(t)\) is the baseline hazard, estimated flexibly using penalised B-splines; \(\text{Comorb}_i\) indicates the presence of any comorbid condition; \(\text{KidneyHist}_i\) denotes a history of other kidney-related issues. The remaining binary covariates encode the use of medications: \(\text{Cortico}_i\) for corticosteroids (prednisolone, methylprednisolone sodium succinate, hydrocortisone, dexamethasone), \(\text{Immuno}_i\) for immunosuppressants (mycophenolate mofetil), \(\text{ImmMod}_i\) for immune modulators (hydroxychloroquine sulfate), \(\text{BCell}_i\) for B-cell depletion therapy (rituximab),
$\text{CCB}_i$ for calcium channel blockers (amlodipine, nifedipine),
and \(\text{ACEi}_i\) for ACE inhibitors (lisinopril, enalapril). Here, \(\eta_i(t)\) is the predicted log-creatinine from the longitudinal sub-model at time \(t\), \(\eta^\prime_i(t)=\frac{d\eta_i(t)}{dt}\) is the instantaneous slope of the biomarker trajectory, \(\int_{0}^{t} \eta_i(s)\, ds\) is the cumulative exposure, and \(\alpha_1, \alpha_2, \alpha_3\) are association parameters linking these longitudinal features to the hazard of the event.

\subsection{Joint Model Specification and Selection}
The model was fitted using the \texttt{JMbayes2} package in \texttt{R} \cite{rizopoulos2016r}. Default priors are weakly informative, providing regularisation while allowing the data to primarily drive inference. Specifically, the fixed-effect coefficients in the longitudinal sub-model, $\beta_0, \dots, \beta_4$, and in the survival sub-model, $\omega_1, \dots, \omega_8$, are assigned Gaussian priors:
\[
\beta_m \sim \mathcal{N}(0, 10^2), \quad m=0,\dots,4, \qquad
\omega_k \sim \mathcal{N}(0, 10^2), \quad k=1,\dots,8.
\]
The variance-covariance matrix of the random effects, $\bm{D}$, is assigned a weakly informative inverse-Wishart prior,
$\bm{D} \sim \text{Inverse-Wishart}(\nu, \Psi)$,
with small degrees of freedom $\nu$ and scale matrix $\Psi$, allowing the data to largely determine the subject-specific random effect variances and covariances.  
\\
The residual variance $\sigma^2$ is assigned a weakly informative inverse-gamma prior, $\sigma^2 \sim \text{Inverse-Gamma}(\nu_\sigma, s_\sigma^2)$, with hyperparameters $\nu_\sigma$ and $s_\sigma^2$ chosen to allow the data to dominate.
Association parameters linking the longitudinal creatinine trajectory to the hazard—$\alpha_1$ (current value), $\alpha_2$ (slope), and $\alpha_3$ (cumulative exposure)—are assigned independent Gaussian priors:
$\alpha_k \sim \mathcal{N}(0, 10^2), \quad k=1,2,3$.\\
The baseline hazard $\lambda_0(t)$ is modelled flexibly with penalised B-splines, using smoothing priors on the spline coefficients to prevent overfitting. Posterior inference is obtained via MCMC, estimating all longitudinal parameters ($\beta_0, \dots, \beta_4$, $b_{0i}, b_{1i}, \sigma^2$), survival parameters ($\omega_1, \dots, \omega_8$), and association parameters ($\alpha_1, \alpha_2, \alpha_3$). \\
We considered seven candidate models (M1-M7). Details of the model selection process are provided in Table~\ref{model_selection_criteria}, which summarises the covariates included in the longitudinal and survival sub-models, together with the corresponding association structures. {For this purpose, all joint models (M1-M7) were estimated using $90000$ MCMC samples, half of which are considered burn-in.}
Convergence was assessed using standard diagnostics, including the Gelman-Rubin $\hat{R}$ statistic and traceplots for all parameters, ensuring proper propagation of uncertainty across both sub-models \cite{gelman1992inference}. \\
Model selection was based on the Watanabe–Akaike Information Criterion (WAIC) \cite{watanabe2010waic} and Log Pseudo-Marginal Likelihood (LPML) \cite{gelfand1994cpo}. Let $\bm{\theta}$ denote the full vector of model parameters and $\mathcal{D}$ the observed longitudinal and survival data, where $\mathcal{D}_i$ denotes the observed data for subject $i$ and $\mathcal{D}_{-i}$ denotes the data from all subjects except subject $i$. The WAIC is defined as $\mathrm{WAIC} = -2(\mathrm{lppd} - p_{\mathrm{WAIC}})$, where $\mathrm{lppd} = \sum_{i=1}^{n}\log E_{\bm{\theta}\mid\mathcal{D}}[p(\mathcal{D}_i\mid\bm{\theta})]$ is the log pointwise predictive density and $p_{\mathrm{WAIC}}$ is the effective number of parameters estimated from the posterior variance of the log-likelihood contributions. The LPML is defined as $\mathrm{LPML} = \sum_{i=1}^{n} \log(\mathrm{CPO}_i)$, where $\mathrm{CPO}_i = p(\mathcal{D}_i \mid \mathcal{D}_{-i})$ is the conditional predictive ordinate for subject $i$. Lower values of WAIC indicate better model fit, whereas higher LPML values indicate better predictive performance.\\
{Among these candidate specifications, model 5 (M5) was selected as the primary model given its superior performance on both WAIC and LPML. In particular, the cumulative exposure association structure provided the best overall model fit among the competing specifications. In the following section, we focus on the results from model M5.} %However, in terms of predictive accuracy, the models using the current value association, as well as the model combining current value and cumulative exposure associations, performed similarly to M5. Therefore, either of these models could also be considered suitable for prediction. 

\subsection{Interpretation of Findings}
We summarise and interpret the posterior estimates of our joint model for both longitudinal creatinine trajectories and the survival outcome.\\
Table \ref{posterior_summary} presents the posterior summaries of the parameters estimated from the fitted joint model M5.
In the longitudinal outcome, all covariates: time, age at entry, BMI z-score and sex are significant. The negative coefficient for sex indicates that females tend to have lower creatinine levels compared to males, while the positive coefficients for time, age, and BMI z-score suggest that creatinine levels increase with increasing values of these covariates.   
\\
%In the survival model, Area(Creatinine) is the association parameter. It has a coefficient of value $2.99$, and a p-value $0$. This suggests that cumulative creatinine levels over time are strongly associated with the risk of developing AKI/CKD. In other words, 
%the risk of diagnosis at time $t$ does not only depend on creatinine level at that time, but on how much the creatinine levels have been high over time. The positive coefficient implies that higher cumulative exposure increases event risk.
%Observe that comorbidity is not significant, which suggests that there is no enough evidence in this dataset to conclude that having a comorbid condition affects the risk of kidney disease. 
%On the other hand, patients with kidney related conditions have a higher risk of being diagnosed with AKI/CKD.
In the survival sub-model, the hazard of the composite kidney-related event (AKI, CKD, or death) is strongly associated with cumulative creatinine exposure over time. The posterior mean for this association parameter is 2.983 (95\% CI: 2.562–3.397, p-value $<$ 0.001), indicating that higher cumulative creatinine levels substantially increase the risk of experiencing the event. Among baseline covariates, a history of kidney-related conditions significantly increases the hazard (posterior mean 0.726, p-value = 0.046), while corticosteroid and calcium channel blocker use are also associated with elevated risk. In contrast, use of immunosuppressants and immune modulators is associated with a reduced hazard. Comorbidity, B-cell therapy, and ACE inhibitor use were not significant predictors of event risk in this cohort. Overall, these results highlight that both the cumulative trajectory of creatinine and specific treatments or prior kidney conditions are key determinants of the risk of AKI, CKD, or death in this population.
\begin{table}[H]
\centering
\footnotesize
\caption{Posterior summaries of joint model parameters for the survival and longitudinal outcomes in the GPKC dataset for model M5. Mean and SD denote the posterior mean and standard deviation, respectively, and 95\% CI denotes 95\% credible intervals.}
\label{posterior_summary}
\begin{tabular}{lcccc}
\toprule
\textbf{Parameter} & \textbf{Mean} & \textbf{SD} & \textbf{95\% CI} & \textbf{p-value} \\
\midrule
\multicolumn{5}{l}{{\textit{Longitudinal Outcome: log(Creatinine)}}} \\
Intercept                  & 3.706   & 0.035  & (3.638, 3.774)       & 0.000 \\
Time                       & 0.063   & 0.004  & (0.055, 0.072)       & 0.000 \\
Sex                        & $-$0.085 & 0.039 & ($-$0.161, $-$0.008) & 0.029 \\
Age at entry (SAge)        & 0.219   & 0.019  & (0.182, 0.257)       & 0.000 \\
BMI z-score (BMIZ)         & 0.044   & 0.009  & (0.027, 0.062)       & 0.000 \\
$\sigma$                   & 0.222   & 0.002  & (0.218, 0.225)       & 0.000 \\
\midrule
\multicolumn{5}{l}{{\textit{Survival Outcome: Hazard of composite kidney-related event}}} \\
Comorbidity (Comorb)                      & 0.087   & 0.534  & ($-$0.998, 1.079)    & 0.848 \\
Kidney condition (KidneyHist)             & 0.726   & 0.357  & (0.012, 1.407)       & 0.046 \\
Corticosteroid (Cortico)                 & 1.043   & 0.394  & (0.307, 1.848)       & 0.006 \\
Immunosuppressant (Immuno)               & $-$1.584 & 0.415 & ($-$2.421, $-$0.788) & 0.000 \\
Immune modulator (ImmMod)                 & $-$1.119 & 0.475 & ($-$2.084, $-$0.226) & 0.013 \\
B-cell therapy (BCell)                    & 0.350   & 0.404  & ($-$0.444, 1.133)    & 0.385 \\
Calcium channel blocker (CCB)             & 1.992   & 0.331  & (1.341, 2.642)       & 0.000 \\
ACE inhibitor (ACEi)                       & 0.163   & 0.424  & ($-$0.683, 0.974)    & 0.693 \\
Cumulative exposure  & 2.983 & 0.213 & (2.562, 3.397)       & 0.000 \\
\bottomrule
\end{tabular}
\end{table}

\subsection{Dynamic Risk Predictions}
\begin{figure}[H]
    \centering
    \captionsetup[subfigure]{font=scriptsize, labelfont=bf, skip=3pt}

    % --- Row 1: Subject 38 ---
    \begin{subfigure}[t]{0.31\textwidth}
        \includegraphics[width=\textwidth, trim=5 5 5 5, clip]{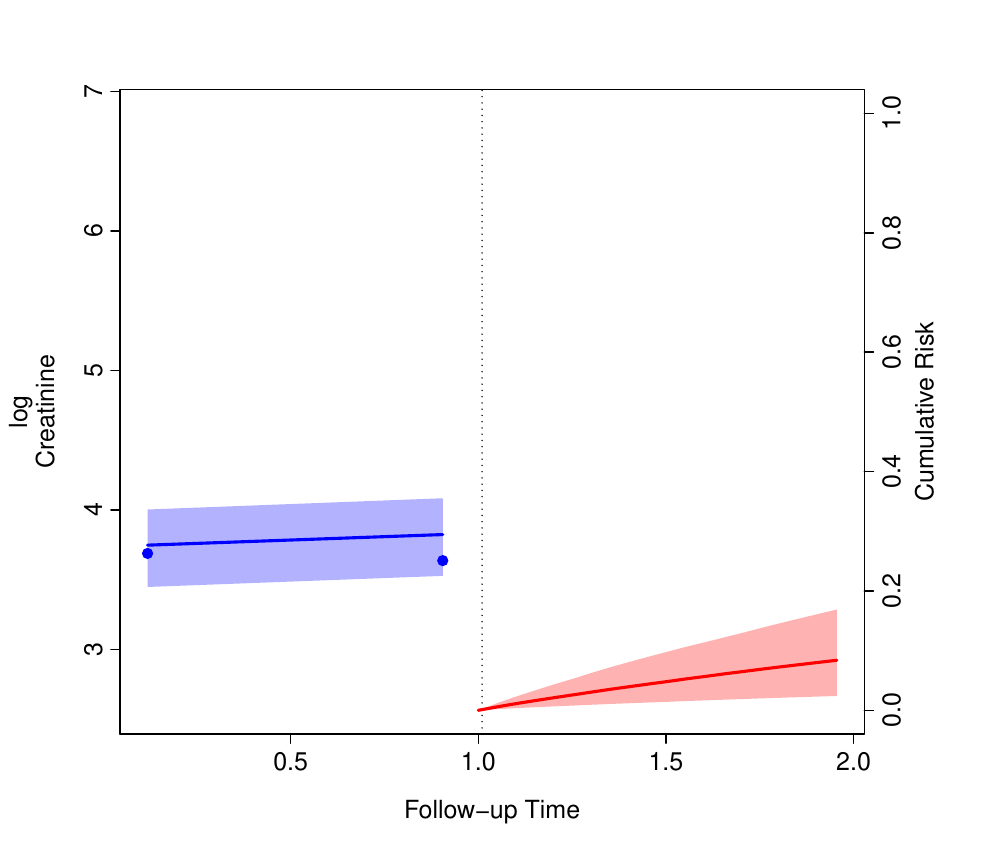}
        \caption{Subject 38, $t_L = 1$}
    \end{subfigure}
    \hfill
    \begin{subfigure}[t]{0.31\textwidth}
        \includegraphics[width=\textwidth, trim=5 5 5 5, clip]{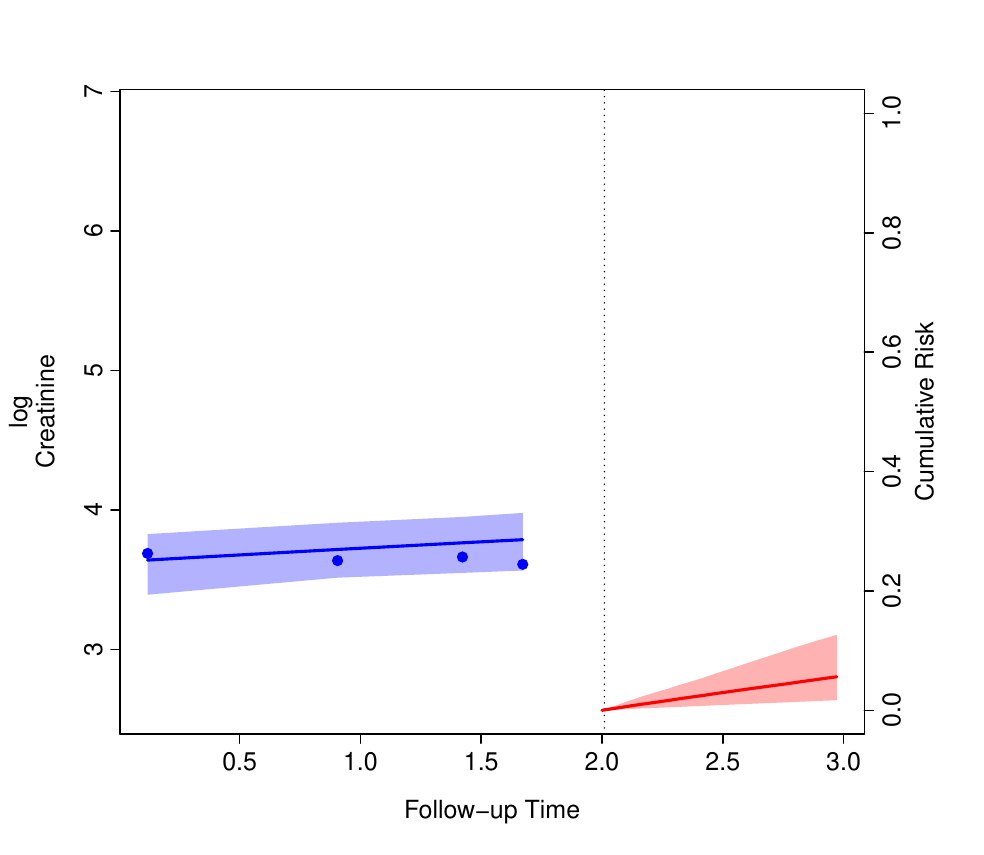}
        \caption{Subject 38, $t_L = 2$ }
    \end{subfigure}
    \hfill
    \begin{subfigure}[t]{0.31\textwidth}
        \includegraphics[width=\textwidth, trim=5 5 5 5, clip]{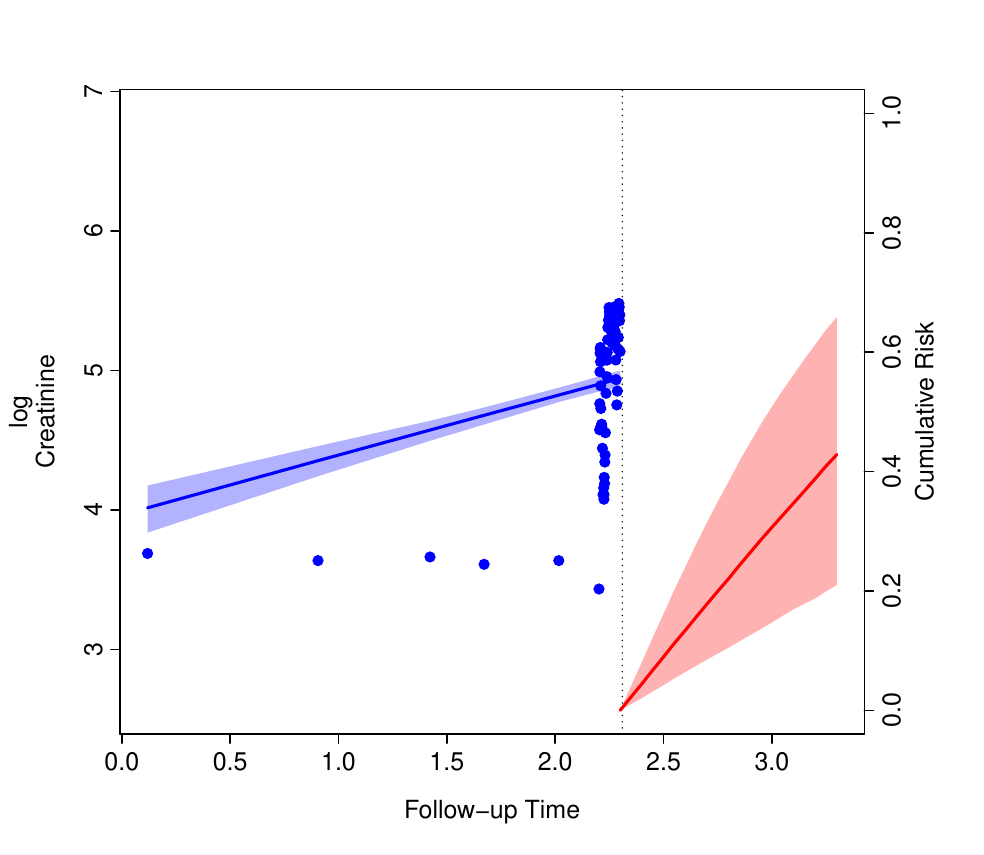}
        \caption{Subject 38, $t_L = 2.3$ }
    \end{subfigure}

    \vspace{4pt}

    % --- Row 2: Subject 313 ---
    \begin{subfigure}[t]{0.31\textwidth}
        \includegraphics[width=\textwidth, trim=5 5 5 5, clip]{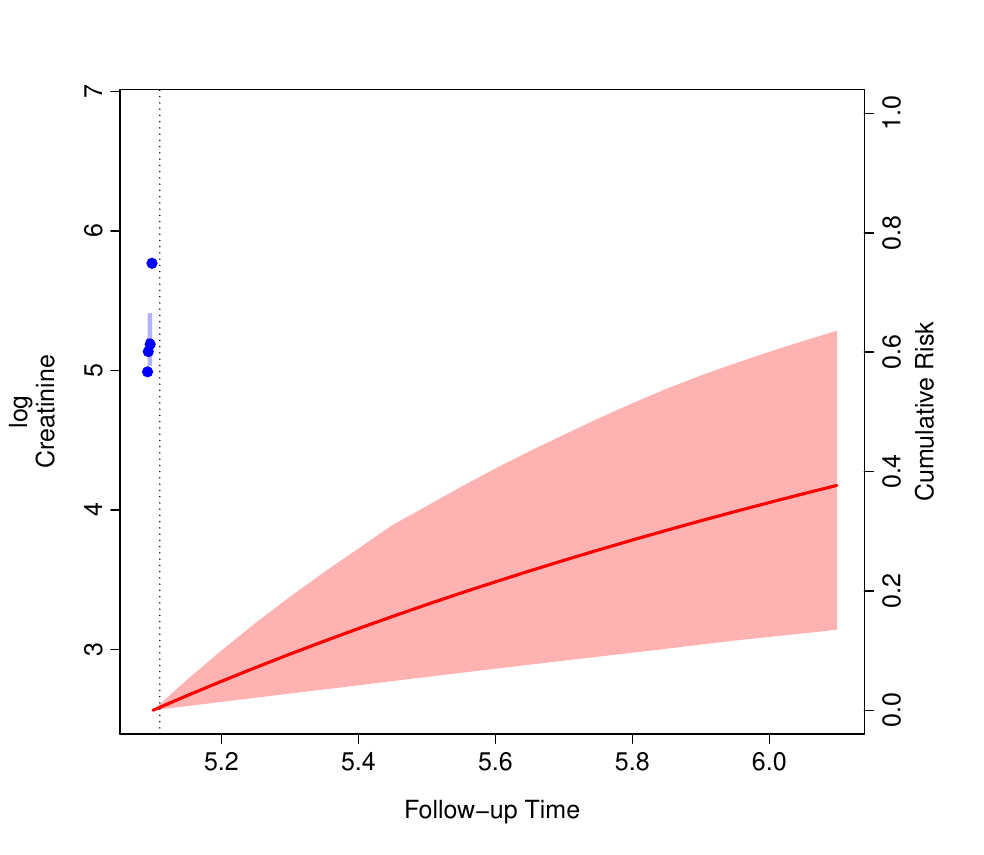}
        \caption{Subject 313, $t_L = 5.1$ }
    \end{subfigure}
    \hfill
    \begin{subfigure}[t]{0.31\textwidth}
        \includegraphics[width=\textwidth, trim=5 5 5 5, clip]{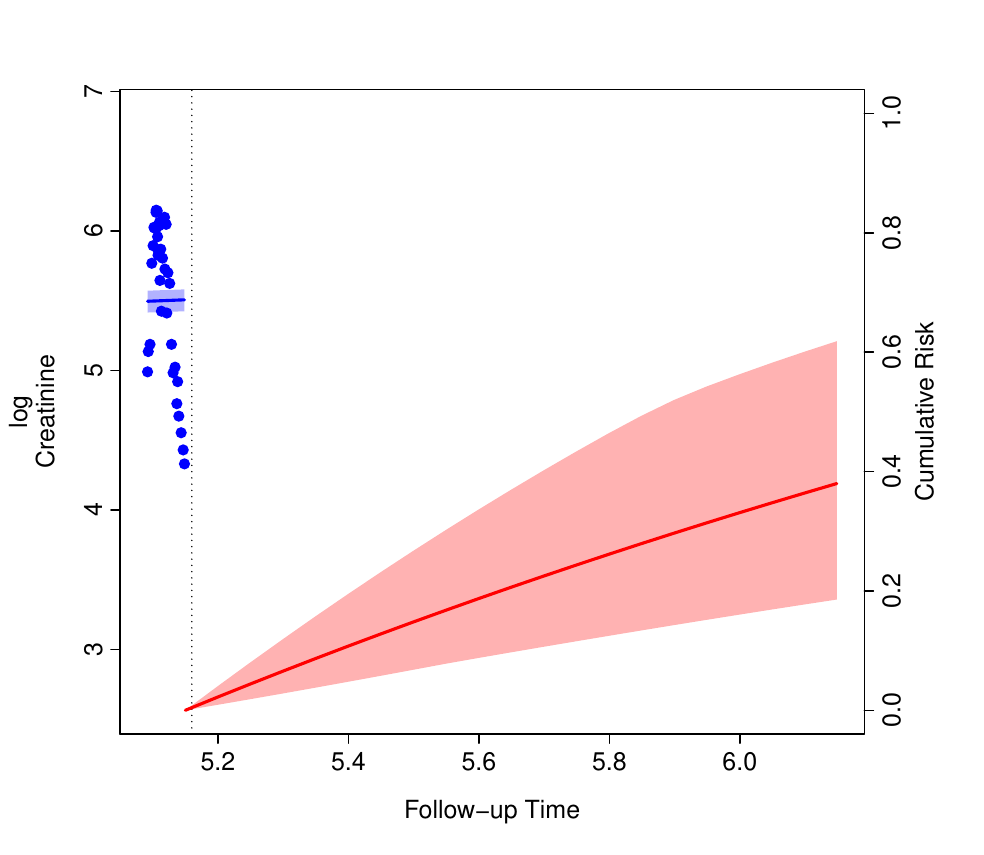}
        \caption{Subject 313, $t_L = 5.15$ }
    \end{subfigure}
    \hfill
    \begin{subfigure}[t]{0.31\textwidth}
        \includegraphics[width=\textwidth, trim=5 5 5 5, clip]{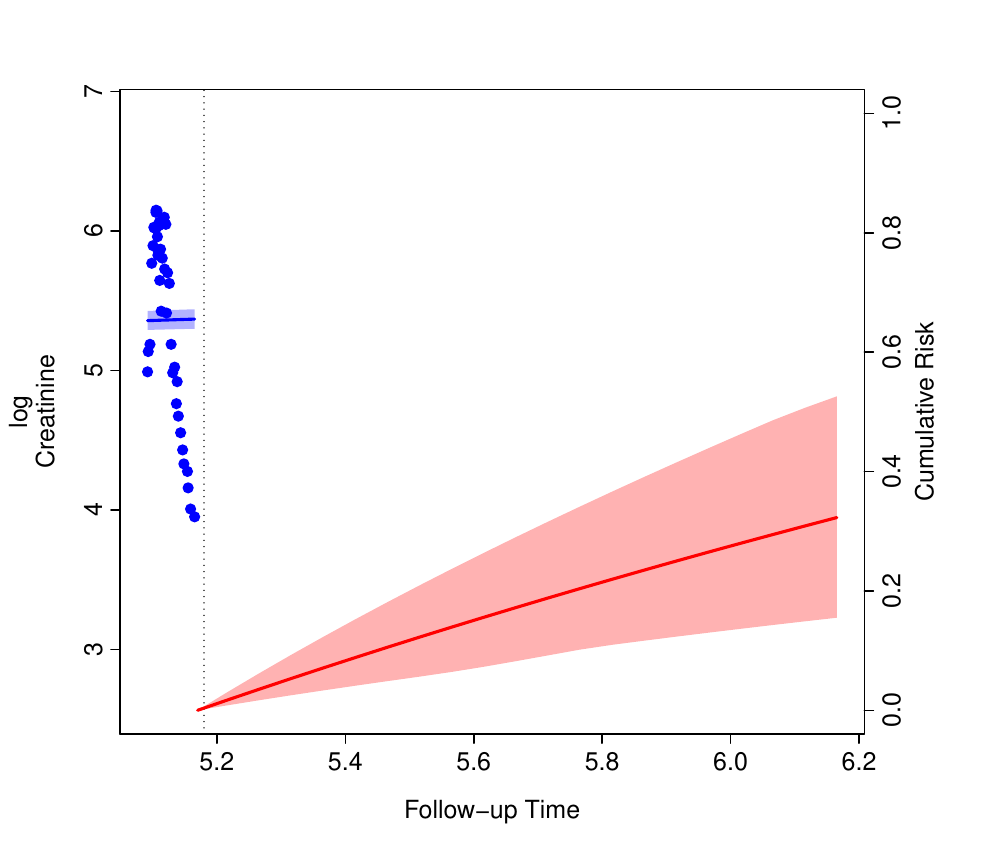}
        \caption{Subject 313, $t_L = 5.17$}
    \end{subfigure}

    \vspace{4pt}

    % --- Row 3: Subject 415 ---
    \begin{subfigure}[t]{0.31\textwidth}
        \includegraphics[width=\textwidth, trim=5 5 5 5, clip]{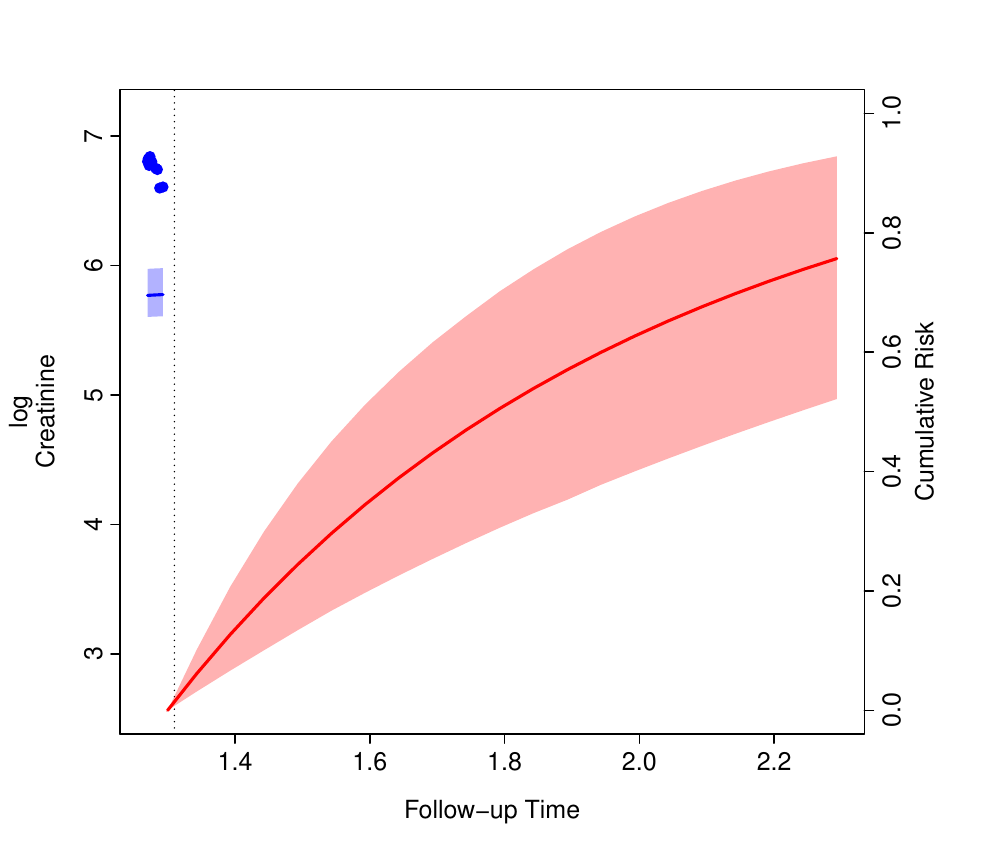}
        \caption{Subject 415, $t_L = 1.3$}
    \end{subfigure}
    \hfill
    \begin{subfigure}[t]{0.31\textwidth}
        \includegraphics[width=\textwidth, trim=5 5 5 5, clip]{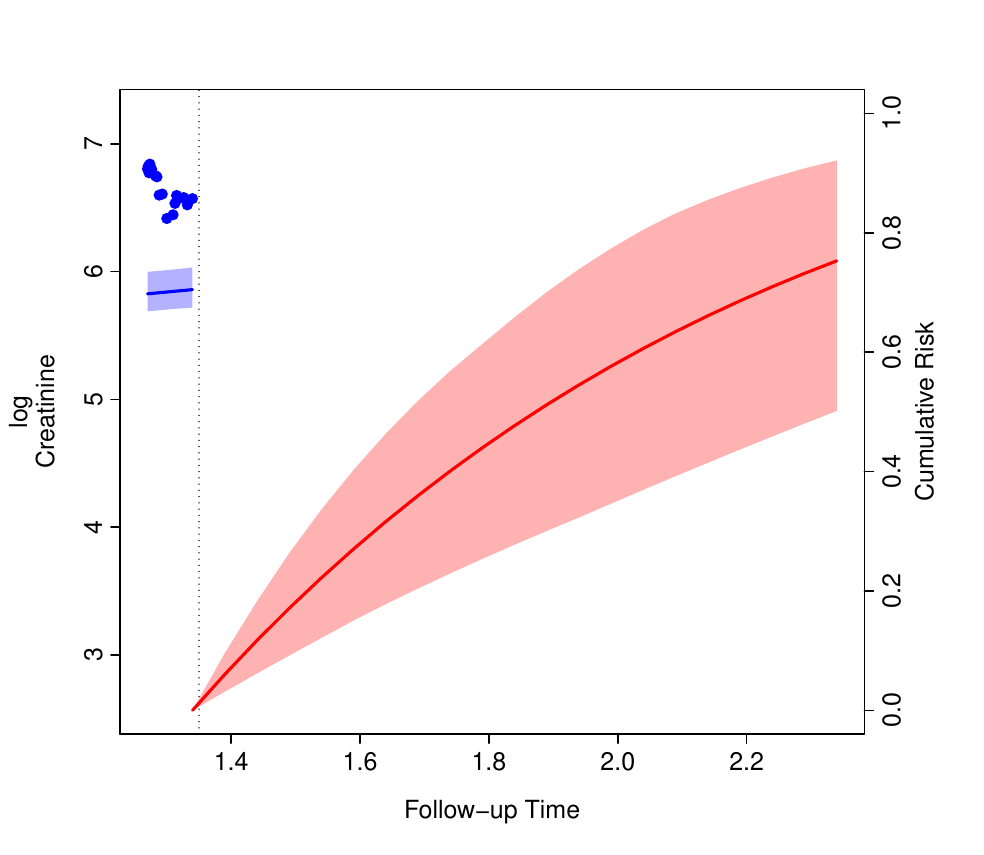}
        \caption{Subject 415, $t_L = 1.34$ }
    \end{subfigure}
    \hfill
    \begin{subfigure}[t]{0.31\textwidth}
        \includegraphics[width=\textwidth, trim=5 5 5 5, clip]{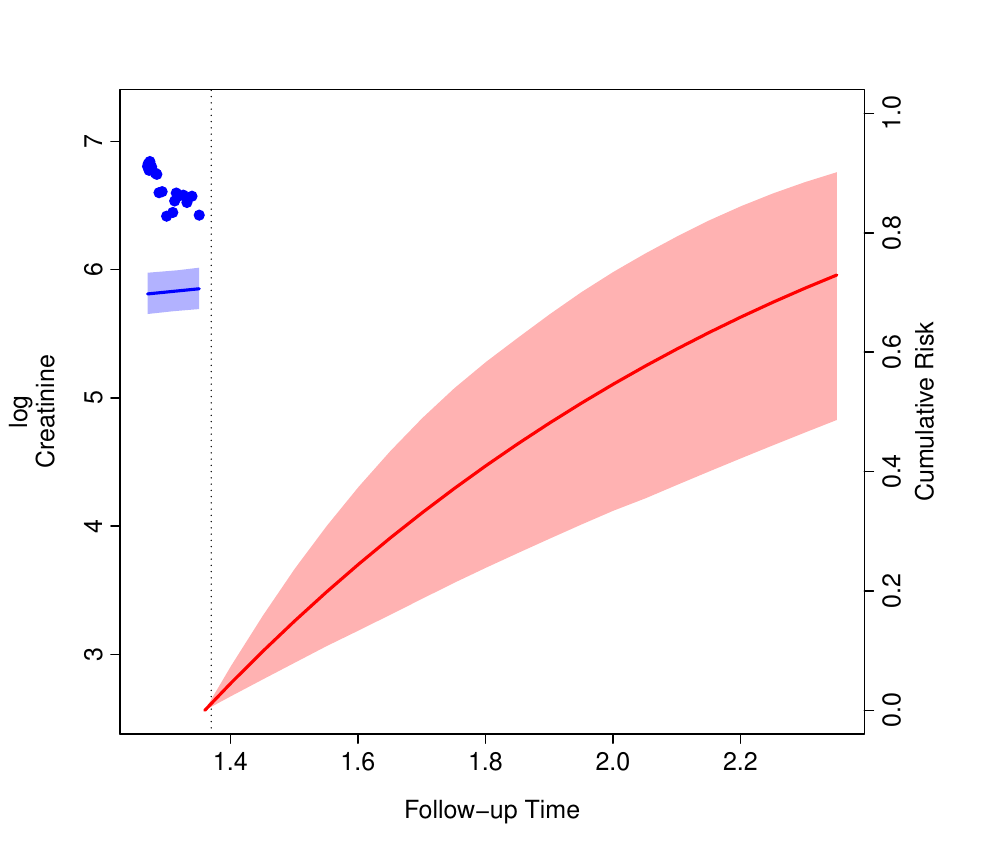}
        \caption{Subject 415, $t_L = 1.36$ }
    \end{subfigure}

    \vspace{6pt}
\caption{Dynamic predictions of survival probability for three patients (38, 313, and 415), evaluated at multiple landmark times $t_L$ with a fixed prediction horizon of $\Delta t = 1$ year. The observed event times are 2.43 years for patient 38, 5.17 years for patient 313, and 1.38 years for patient 415.}
    \label{dynamic_predictions}
\end{figure}

%Dynamic predictions allow us to forecast both future creatinine levels and the risk of experiencing the composite event at any time up to the event occurrence. These predictions can be updated whenever a new creatinine measurement becomes available, hence the term dynamic predictions \cite{rizopoulos_2010,hashemi2025dynamic,ganjali2024joint}.\\
%Consider a scenario in which three creatinine measurements have been collected. We aim to predict future creatinine values and the probability of experiencing the event up to a future time $t$. Two key concepts in dynamic prediction are:
% {Landmark time ($t_L$)}: the time up to which longitudinal measurements have been observed.
%{Horizon time ($t_L + \Delta t$)}: the future interval over which the predictions are made.\\
%To evaluate the predictive performance of the model, we used two commonly employed metrics: the time-dependent area under the ROC curve (AUC) and the time-dependent Brier score. The AUC measures the model’s ability to discriminate between patients who experience the event and those who do not. The Brier score quantifies prediction error, indicating the accuracy of the predicted event probabilities.\\
For dynamic prediction, we evaluated 4-fold cross-validated predictive performance using Model 5, which was selected as the best-fitting model. \\
We considered landmark times $t_L = 0.5$ and $2$ years, and prediction windows $\Delta t = 1, 2, 3$ years, in order to assess short- and medium-term risk predictions.\\
Table~\ref{auc_brier} summarises the predictive performance by landmark time and prediction window, based on AUC and BS. Overall, AUC values are relatively high, with a slight decline as the prediction window increases. BS values remain low, indicating accurate risk predictions. Nevertheless, these results should be interpreted with caution due to the high censoring rate (89\%).\\
Figure \ref{dynamic_predictions} displays dynamic risk predictions along with the corresponding 95\% credible intervals for three randomly selected patients in our cohort (38, 313, 415) who were at risk at these time points. The blue lines indicate the predicted creatinine trajectories, while the red curves represent the predicted probability of experiencing the event. Predictions were generated at different landmark times with a fixed prediction horizon of $\Delta t = 1$ year, corresponding to the most clinically relevant window for this cohort.
\\
For patient 38, at landmark time $t_L=1$ year, the model incorporates all creatinine measurements collected up to $t=1$ and predicts a relatively low risk of event over the subsequent year (panel a). As additional measurements become available at $t_L=2$, the creatinine values are lower than previously observed, suggesting a slight improvement in renal function, and the model responds by reducing the predicted event risk (panel b). By $t_L=2.3$, however, a sharp rise in creatinine is recorded, and the model adjusts accordingly, yielding a substantially higher predicted event probability (panel c).
\\
Patient 313 presents with high log-creatinine values and measurements taken at closely spaced time intervals, indicative of intensive clinical monitoring. Despite the elevated baseline, creatinine levels decline rapidly over a short period, and the model captures this improvement through a decreasing predicted event probability across landmark times (panels d-f).  
In contrast, patient 415 exhibits persistently elevated log-creatinine levels throughout the observation period. Consistent with this trajectory, the model assigns a high and stable predicted event probability across all landmark times (panels g-i).
Overall, these results demonstrate that predictions are updated dynamically as new longitudinal measurements are incorporated, and that the predicted survival trajectories differ markedly across subjects. This highlights the joint model's capacity for personalised risk prediction, tailoring survival estimates to each patient's individual longitudinal history.

\begin{table}[ht]
\centering
\footnotesize
\caption{Predictive performance of model M5 for the GPKC study: time-dependent area under the ROC curve (AUC) and Brier score (BS) evaluated at each landmark time $t_L$ and prediction window $\Delta t$ (in years).}\label{model5_performance}
\begin{tabular}{lcccccc}
\toprule
& \multicolumn{3}{c}{\textbf{AUC}} & \multicolumn{3}{c}{\textbf{BS}} \\
\cmidrule(lr){2-4} \cmidrule(lr){5-7}
\textbf{$t_L$} & $\Delta t = 1$ & $\Delta t = 2$ & $\Delta t = 3$ & $\Delta t = 1$ & $\Delta t = 2$ & $\Delta t = 3$ \\
\midrule
0.5 & 0.803 (0.205) & 0.871 (0.078) & 0.868 (0.026) & 0.0216 (0.0042) & 0.0223 (0.0058) & 0.0243 (0.0045) \\
2.0 & 0.784 (0.168) & 0.807 (0.216) & 0.764 (0.214) & 0.0283 (0.0181) & $\dag$ & 0.0282 (0.0190) \\
\bottomrule
\end{tabular}
\medskip
\begin{minipage}{0.92\textwidth}
\small
\textit{$\dag$:} The Brier Score could not be computed for this point using \texttt{JMbayes2}.\end{minipage}
\label{auc_brier}
\end{table}

\section{Discussion}\label{sec12}
In this paper, we applied a Bayesian joint model for longitudinal and survival data to the GPKC cohort, focusing on paediatric patients. While several studies have examined factors related to kidney disease risk \cite{de2003predicting, liao2024approach}, most focus on adult populations or rely on survival models with time-varying covariates \cite{atkinson2021ckid, ng2019incidence, weidemann2020plasma}. In contrast, our analysis simultaneously models longitudinal creatinine measurements, a widely used biomarker of kidney function, and time to diagnosis of AKI/CKD, enabling a detailed assessment of their association in children.\\
Joint modelling provides a statistically rigorous framework that links longitudinal biomarker trajectories with time-to-event outcomes. Unlike separate analyses, it accounts for measurement error in the biomarker, potential informative dropout, and the temporal correlation between repeated measurements and the event process. This allows for more accurate estimation of biomarker effects on event risk and facilitates individualised, time-updated risk predictions. Additionally, joint modelling easily accommodates right censoring and late entry.\\
Joint models have been used to study kidney disease progression in adults \cite{liao2024approach, stamenicprognostic, belay2023joint}; however, these studies focus predominantly on time to ESKD in patients already diagnosed with CKD. Our study instead targets earlier disease stages, with time-to-event defined as time to first diagnosis of AKI or CKD or death, aiming to identify factors predictive of disease onset.\\
While a few studies use eGFR as the longitudinal biomarker \cite{liao2024approach}, we model serum creatinine directly, as it is a by-product of muscle metabolism excreted solely via the kidneys, making it a excellent biomarker of renal function.
Consistent with previous findings \cite{de_bruijne_et_al_2003, stamenicprognostic, belay_et_al_2023}, our results suggest that higher serum creatinine levels are associated with worse renal outcomes. 
Our analysis indicates that cumulative exposure to elevated creatinine levels is strongly associated with an increased risk of AKI/CKD. 
This highlights the importance of jointly modelling longitudinal biomarkers and time-to-event outcomes, and in particular the cumulative exposure association structure, as the risk is influenced not only by the current creatinine level but also by the history of biomarker elevations over time.
Moreover, unlike \cite{stamenicprognostic} who model the association between serum creatinine and risk of graft failure via latent classes, our association structure allows for direct quantification of the effect of serum creatinine on the time to AKI/CKD diagnosis.\\
The longitudinal serum creatinine measurements were highly skewed, so we applied a log-transformation, which substantially reduced skewness. Although this transformation improved symmetry, some residual asymmetry remains. For future work, further exploration of data normalisation techniques, alternative transformation methods, or other distributional assumptions could be considered to address this remaining skewness \cite{baghfalaki2014bayesian,baghfalaki2015bayesian}.\\
Regarding medication effects, corticosteroid use was associated with an increased hazard of AKI/CKD, consistent with prior paediatric and adult studies \cite{friedman1982glucocorticoids, ponticelli2018glucocorticoids}. Treatment duration, timing, and individual renal conditions can modulate corticosteroid effects, and our analysis did not account for these factors. Conversely, immunosuppressant use was associated with reduced risk, aligning with evidence supporting their benefit in treating paediatric renal conditions \cite{faedda1995immunosuppressive, esdaile1994benefit}. Calcium channel blockers (CCBs) were associated with higher risk of AKI/CKD, consistent with studies suggesting that certain CCBs, such as nifedipine, can increase glomerular pressure, potentially adversely affecting kidney function \cite{wolf2004all, hadtstein2008hypertension}. Contrary to previous findings \cite{belay2023joint, atkinson2021ckid}, comorbid conditions did not show a significant association with event risk in this cohort, suggesting limited predictive value in this population.\\
A primary objective of studies on renal disease risk is providing individualised risk predictions to support clinical decision-making. 
A key advantage of the joint modelling framework is dynamic prediction. Whereas  \cite{atkinson2021ckid} adapted a clinical tool for predicting CKD progression risk in the CKiD study using baseline covariates, our model provides individualised, dynamic risk predictions that incorporate all available creatinine measurements, offering a practical tool for early identification of high-risk patients. These predictions update once new creatinine measurements become available. Traditional survival models with time-varying covariates often treat biomarker measurements as fixed and ignore measurement error, potentially biasing risk estimates. By jointly modelling the longitudinal and survival processes, we capture both the magnitude and cumulative history of biomarker changes, improving risk stratification. The strong AUC and BS obtained attest to the reliability of the dynamic prediction tool in providing accurate individualised risk predictions for AKI/CKD diagnosis.\\
Overall, these findings highlight the clinical utility of joint models in paediatric nephrology. By simultaneously modelling longitudinal biomarkers and time-to-event outcomes, joint models provide enhanced risk assessment compared to conventional survival analysis, offering both improved predictive accuracy and insights into how biomarker trajectories influence event risk over time.\\
Future studies could extend this work by validating the joint modelling framework in larger or multi-center paediatric cohorts and incorporating additional biomarkers of kidney function to enhance predictive accuracy. Model extensions could include nonlinear longitudinal trajectories, time-varying treatment effects, or competing risk models. Furthermore, translating dynamic risk predictions into real-time clinical decision-support tools could facilitate early intervention in high-risk patients. Finally, more sophisticated methods to handle missing covariate data not missing at random (MNAR) could improve robustness of the model when the MAR assumption may not hold.

\subsection*{Acknowledgements}
This work was facilitated by the Great Ormond Street Hospital for Children NHS Foundation Trust (GOSH) Digital Research Environment (DRE).

\bibliographystyle{plain}
\bibliography{wileyNJD-AMA.bib}

@article{albert2010estimating,
  title={On estimating the relationship between longitudinal measurements and time-to-event data using a simple two-stage procedure},
  author={Albert, Paul S and Shih, Joanna H},
  journal={Biometrics},
  volume={66},
  number={3},
  pages={983--987},
  year={2010},
  publisher={Wiley Online Library}
}

@article{laird1982random,
  title={Random-effects models for longitudinal data},
  author={Laird, Nan M and Ware, James H},
  journal={Biometrics},
  volume={38},
  number={4},
  pages={963--974},
  year={1982},
  publisher={JSTOR}
}

@article{tsiatis2004joint,
  title={Joint modeling of longitudinal and time-to-event data: an overview},
  author={Tsiatis, Anastasios A and Davidian, Marie},
  journal={Statistica Sinica},
  pages={809--834},
  year={2004},
  publisher={JSTOR}
}

@article{rizopoulos2011dynamic,
  title={Dynamic predictions and prospective accuracy in joint models for longitudinal and time-to-event data},
  author={Rizopoulos, Dimitris},
  journal={Biometrics},
  volume={67},
  number={3},
  pages={819--829},
  year={2011},
  publisher={Wiley},
  doi={10.1111/j.1541-0420.2010.01546.x}
}

@article{taylor2013real,
  title={Real-time individual predictions of prostate cancer recurrence using joint models},
  author={Taylor, J.M.G. and Park, Y. and others},
  journal={Biometrics},
  volume={69},
  number={1},
  pages={206--214},
  year={2013},
  doi={10.1111/biom.12024}
}

@article{rizopoulos2014combining,
  title={Combining dynamic predictions from joint models for longitudinal and time-to-event data using Bayesian model averaging},
  author={Rizopoulos, Dimitris and Hatfield, Laura A. and Carlin, Bradley P. and Takkenberg, Johanna J. M.},
  journal={Journal of the American Statistical Association},
  volume={109},
  number={508},
  pages={1385--1397},
  year={2014},
  doi={10.1080/01621459.2014.931236}
}

@book{daniels2008missing,
  title={Missing Data in Longitudinal Studies: Strategies for Bayesian Modeling and Sensitivity Analysis},
  author={Daniels, Michael J. and Hogan, Joseph W.},
  year={2008},
  publisher={CRC Press}
}

@article{gelman1992inference,
  title={Inference from Iterative Simulation Using Multiple Sequences},
  author={Gelman, Andrew and Rubin, Donald B.},
  journal={Statistical Science},
  volume={7},
  number={4},
  pages={457--472},
  year={1992},
  publisher={Institute of Mathematical Statistics}
}

@article{watanabe2010waic,
  title={Asymptotic equivalence of Bayes cross validation and widely applicable information criterion in singular learning theory},
  author={Watanabe, Sumio},
  journal={Journal of Machine Learning Research},
  volume={11},
  pages={3571--3594},
  year={2010}
}

@article{gelfand1994cpo,
  title={Model determination using predictive distributions with implementation via sampling-based methods},
  author={Gelfand, Alan E and Dey, Dipak K},
  journal={Bayesian Statistics},
  volume={4},
  pages={147--167},
  year={1994}
}

@article{baghfalaki2015bayesian,
  title={A Bayesian approach for joint modeling of skew-normal longitudinal measurements and time to event data},
  author={Baghfalaki, Taban and Ganjali, Mojtaba},
  journal={REVSTAT-Statistical Journal},
  volume={13},
  number={2},
  pages={169--191},
  year={2015}
}

@article{rizopoulos2016r,
  title={The R package JMbayes for fitting joint models for longitudinal and time-to-event data using MCMC},
  author={Rizopoulos, Dimitris},
  journal={Journal of statistical software},
  volume={72},
  pages={1--46},
  year={2016}
}

@article{wulfsohn1997,
  title={A joint model for survival and longitudinal data measured with error},
  author={Wulfsohn, M. S. and Tsiatis, A. A.},
  journal={Biometrics},
  volume={53},
  number={1},
  pages={330--339},
  year={1997}
}

@article{tsiatis2004,
  title={Joint modeling of longitudinal and time-to-event data: An overview},
  author={Tsiatis, A. A. and Davidian, M.},
  journal={Statistica Sinica},
  volume={14},
  pages={809--834},
  year={2004}
}

@article{liao2024approach,
  title={An approach for personalized dynamic assessment of chronic kidney disease progression using joint model},
  author={Liao, Chen-Mao and Kao, Yi-Wei and Chang, Yi-Ping and Lin, Chih-Ming},
  journal={Biomedicines},
  volume={12},
  number={3},
  pages={622},
  year={2024},
  publisher={MDPI}
}

@article{armero2018,
  title={Bayesian joint modeling for assessing the progression of chronic kidney disease in children},
  author={Armero, C. and Forte, A. and Perpi{\~n}{\'a}n, H. and Sanahuja, M. J. and Agust{\'\i}, S.},
  journal={Statistical Methods in Medical Research},
  volume={27},
  number={1},
  pages={298--311},
  year={2018},
  doi={10.1177/0962280216628560}
}

@article{sayyadi2017assessing,
  title={Assessing risk indicators of allograft survival of renal transplant: An application of joint modeling of longitudinal and time-to-event analysis},
  author={Sayyadi, Hojjat and Zayeri, Farid and Baghestani, Ahmad Reza and Baghfalaki, Taban and Afshari, Ali Taghizadeh and Mohammadrahimi, Mohsen and Fereidoni, Javid and Makhdoomi, Khadijeh},
  journal={Iran Red Crescent Med J},
  volume={19},
  number={3},
  pages={e40583},
  year={2017}
}

@book{verbeke2000,
  title={Linear Mixed Models for Longitudinal Data},
  author={Verbeke, G. and Molenberghs, G.},
  publisher={Springer},
  year={2000}
}

@article{de_bruijne_et_al_2003,
title = {Predicting kidney graft failure using time-dependent renal function covariates},
journal = {Journal of Clinical Epidemiology},
volume = {56},
number = {5},
pages = {448-455},
year = {2003},
issn = {0895-4356},
doi = {https://doi.org/10.1016/S0895-4356(03)00004-0},
url = {https://www.sciencedirect.com/science/article/pii/S0895435603000040},
author = {de Bruijne, Mattheus H.J and Sijpkens, Yvo W.J and Paul, Leendert C. and Westendorp, Rudi G.J and van Houwelingen, Hans C. and Zwinderman, Aeilko H. }
}

@article{blanche_et_al_2015,
title = {Quantifying and comparing dynamic predictive accuracy of joint models for longitudinal marker and time-to-event in presence of censoring and competing risks},
journal = {Biometrics},
volume = {71},
pages = {102-113},
year = {2015},
doi = { https://doi.org/10.1111/biom.12232},
author = {Blanche, Paul and Proust-Lima, Cecile and Loubere, Lucie and Berr, Claudine and Dartigues, Jean-Francois and Jacqmin-Gadda, Helene }
}

@article{papageorgiou_et_al_2019,
author = {Papageorgiou, Grigorios and Mauff, Katya and Tomer, Anirudh and Rizopoulos, Dimitris },
title ={An Overview of Joint Modeling of Time-to-Event and Longitudinal Outcomes},
journal = {Annual Review of Statistics and Its Application},
volume = {6},
issue = {1},
pages = {223-240},
year = {2019},
URL = {http://dx.doi.org/10.1146/annurev-statistics-030718-105048},
SSRN = {https://ssrn.com/abstract=3382097}
}

@article{cox_1972,
 ISSN = {00359246},
 URL = {http://www.jstor.org/stable/2985181},
 author = {D. R. Cox},
 journal = {Journal of the Royal Statistical Society. Series B (Methodological)},
 number = {2},
 pages = {187--220},
 publisher = {[Royal Statistical Society, Wiley]},
 title = {Regression Models and Life-Tables},
 urldate = {2024-05-24},
 volume = {34},
 year = {1972}
}

@article{jager_et_al_2019,
title = {A single number for advocacy and communication—worldwide more than 850 million individuals have kidney diseases},
journal = {Kidney International},
volume = {96},
number = {5},
pages = {1048-1050},
year = {2019},
issn = {0085-2538},
doi = {https://doi.org/10.1016/j.kint.2019.07.012},
url = {https://www.sciencedirect.com/science/article/pii/S0085253819307860},
author = {Kitty J. Jager and Csaba Kovesdy and Robyn Langham and Mark Rosenberg and Vivekanand Jha and Carmine Zoccali}
}

@article{saran_et_al_2020,
title = {US Renal Data System 2019 Annual Data Report: Epidemiology of Kidney Disease in the United States},
journal = {American Journal of Kidney Diseases},
volume = {75},
number = {1, Supplement 1},
pages = {A6-A7},
year = {2020},
note = {US Renal Data System 2019 Annual Data Report},
issn = {0272-6386},
doi = {https://doi.org/10.1053/j.ajkd.2019.09.003},
url = {https://www.sciencedirect.com/science/article/pii/S0272638619310091},
author = {Saran, Rajiv  and Robinson, Bruce  and Abbott, Kevin C.  and Bragg-Gresham et al., Jennifer  }
}

@article{kovesdy_2022,
title = {Epidemiology of chronic kidney disease: an update 2022},
journal = {Kidney International Supplements},
volume = {12},
number = {1},
pages = {7-11},
year = {2022},
note = {Aldosterone and the Mineralocorticoid Receptor: Emerging Therapeutic Paradigms for Cardiorenal Protection},
issn = {2157-1716},
doi = {https://doi.org/10.1016/j.kisu.2021.11.003},
url = {https://www.sciencedirect.com/science/article/pii/S2157171621000666},
author = {Csaba P. Kovesdy}
}

@article{belay_et_al_2023,
    author = {Belay, Samson and Melese, Dessie and Muhammed, Kasim},
    title = "{Joint modeling on serum creatinine and time to end stage of renal disease for chronic kidney disease patients under treatment at the University of Gondar Referral Hospital}",
    journal = {Health Science Reports},
    pages = {e1563},
    year = {2023},
    doi = {doi:10.1002/hsr2.1563},
    url = {https://doi.org/10.1002/hsr2.1563},
    eprint = {https://onlinelibrary.wiley.com/doi/full/10.1002/hsr2.1563},
}

@article{bibkov_et_al_2019,
  title={Global, regional, and national burden of chronic kidney disease, 1990--2017: a systematic analysis for the Global Burden of Disease Study 2017},
  author={Bikbov, Boris and Purcell, Caroline A and Levey, Andrew S and Smith, Mari and Abdoli, Amir and Abebe, Molla and Adebayo, Oladimeji M and Afarideh, Mohsen and Agarwal, Sanjay Kumar and Agudelo-Botero, Marcela and others},
  journal={The lancet},
  volume={395},
  number={10225},
  pages={709--733},
  year={2020},
  publisher={Elsevier}
}

@article{children_kidney, 
    author = {Kula, Alexander J.1; Somers, Michael J.G.2;  on behalf of the American Society of Pediatric},
    title = "{Children with CKD Are Not Little Adults with CKD: Pediatric Considerations for the Advancing American Kidney Health Initiative}",
    journal = {CJASN},
    year = {2021},
    month = {March},
    issue = {16}, 
    volume = {3}, 
    pages = {470-472}, 
    doi = {10.2215/CJN.11540720}
}

@article{cole1998british,
  title={British 1990 growth reference centiles for weight, height, body mass index and head circumference fitted by maximum penalized likelihood},
  author={Cole, Tim J and Freeman, Jenny V and Preece, Michael A},
  journal={Statistics in medicine},
  volume={17},
  number={4},
  pages={407--429},
  year={1998},
  publisher={Wiley Online Library}
  }

@book{rubin1987multiple,
  title={Multiple Imputation for Nonresponse in Surveys},
  author={Rubin, Donald B},
  year={1987},
  publisher={John Wiley \& Sons},
  address={New York, NY}
}

@article{baghfalaki2014bayesian,
  title={Bayesian joint modeling of longitudinal measurements and time-to-event data using robust distributions},
  author={Baghfalaki, Taban and Ganjali, Mojtaba and Hashemi, Reza},
  journal={Journal of biopharmaceutical statistics},
  volume={24},
  number={4},
  pages={834--855},
  year={2014},
  publisher={Taylor \& Francis}
}

@article{guo2004separate,
  title={Separate and joint modeling of longitudinal and event time data using standard computer packages},
  author={Guo, Xu and Carlin, Bradley P},
  journal={The american statistician},
  volume={58},
  number={1},
  pages={16--24},
  year={2004},
  publisher={Taylor \& Francis}
}

@book{rizopoulos2012joint,
  title={Joint Models for Longitudinal and Time-to-Event Data: With Applications in R},
  author={Rizopoulos, Dimitris},
  year={2012},
  publisher={CRC Press},
  address={Boca Raton, FL}
}

@Manual{jmbayes2,
  title = {JMbayes2: Extended Joint Models for Longitudinal and Time-to-Event Data},
  author = {Dimitris Rizopoulos and Pedro Miranda-Afonso and Grigorios Papageorgiou},
  year = {2026},
  note = {R package version 0.6-0},
  url = {https://drizopoulos.github.io/JMbayes2/},
}

@article{rizopoulos2017dynamic,
  title={Dynamic predictions with time-dependent covariates in survival analysis using joint modeling and landmarking},
  author={Rizopoulos, Dimitris and Molenberghs, Geert and Lesaffre, Emmanuel MEH},
  journal={Biometrical Journal},
  volume={59},
  number={6},
  pages={1261--1276},
  year={2017},
  publisher={Wiley Online Library}
}

@article{atkinson2021ckid,
  title={The CKiD study: overview and summary of findings related to kidney disease progression},
  author={Atkinson, Meredith A and Ng, Derek K and Warady, Bradley A and Furth, Susan L and Flynn, Joseph T},
  journal={Pediatric Nephrology},
  volume={36},
  number={3},
  pages={527--538},
  year={2021},
  publisher={Springer}
}

@article{ng2019incidence,
  title={Incidence of initial renal replacement therapy over the course of kidney disease in children},
  author={Ng, Derek K and Matheson, Matthew B and Warady, Bradley A and Mendley, Susan R and Furth, Susan L and Mu{\~n}oz, Alvaro},
  journal={American journal of epidemiology},
  volume={188},
  number={12},
  pages={2156--2164},
  year={2019},
  publisher={Oxford University Press}
}

@article{weidemann2020plasma,
  title={Plasma soluble urokinase plasminogen activator receptor (suPAR) and CKD progression in children},
  author={Weidemann, Darcy K and Abraham, Alison G and Roem, Jennifer L and Furth, Susan L and Warady, Bradley A},
  journal={American Journal of Kidney Diseases},
  volume={76},
  number={2},
  pages={194--202},
  year={2020},
  publisher={Elsevier}
}

@book{little2019statistical,
  title={Statistical analysis with missing data},
  author={Little, Roderick JA and Rubin, Donald B},
  year={2019},
  publisher={John Wiley \& Sons}
}

@Manual{childsds,
    title = {childsds: Data and Methods Around Reference Values in
      Pediatrics},
    author = {Mandy Vogel},
    year = {2025},
    note = {R package version 0.9.11},
    url = {https://CRAN.R-project.org/package=childsds},
  }

@article{de2003predicting,
  title={Predicting kidney graft failure using time-dependent renal function covariates},
  author={de Bruijne, Mattheus HJ and Sijpkens, Yvo WJ and Paul, Leendert C and Westendorp, Rudi GJ and van Houwelingen, Hans C and Zwinderman, Aeilko H},
  journal={Journal of clinical epidemiology},
  volume={56},
  number={5},
  pages={448--455},
  year={2003},
  publisher={Elsevier}
}

@article{ponticelli2018glucocorticoids,
  title={Glucocorticoids in the treatment of glomerular diseases: pitfalls and pearls},
  author={Ponticelli, Claudio and Locatelli, Francesco},
  journal={Clinical Journal of the American Society of Nephrology},
  volume={13},
  number={5},
  pages={815--822},
  year={2018},
  publisher={LWW}
}

@article{friedman1982glucocorticoids,
  title={Glucocorticoids in renal disease: Theoretical basis, consequences and efficacy of use in the pediatric patient},
  author={Friedman, Aaron L and Chesney, Russell W},
  journal={American journal of nephrology},
  volume={2},
  number={6},
  pages={330--341},
  year={1982},
  publisher={S. Karger AG Basel, Switzerland}
}

@article{faedda1995immunosuppressive,
  title={Immunosuppressive treatment of the glomerulonephritis of systemic lupus.},
  author={Faedda, R and Palomba, D and Satta, A and Pirisi, Mario and Tanda, F and Bartoli, E},
  journal={Clinical nephrology},
  volume={44},
  number={6},
  pages={367--375},
  year={1995}
}

@article{esdaile1994benefit,
  title={The benefit of early treatment with immunosuppressive agents in lupus nephritis.},
  author={Esdaile, JM and Joseph, L and MacKenzie, T and Kashgarian, M and Hayslett, JP},
  journal={The Journal of Rheumatology},
  volume={21},
  number={11},
  pages={2046--2051},
  year={1994}
}

@article{wolf2004all,
  title={Are all antihypertensive drugs renoprotective?},
  author={Wolf, Sabine and Risler, Teut},
  journal={Herz},
  volume={29},
  number={3},
  pages={248},
  year={2004},
  publisher={Springer Nature BV}
}

@article{hadtstein2008hypertension,
  title={Hypertension in children with chronic kidney disease: pathophysiology and management},
  author={Hadtstein, Charlotte and Schaefer, Franz},
  journal={Pediatric Nephrology},
  volume={23},
  number={3},
  pages={363--371},
  year={2008},
  publisher={Springer}
}

@misc{stamenicprognostic,
  title={A prognostic tool for individualized prediction of graft failure risk within ten years after kidney transplantation. J Transplant. 2019; 2019: 7245142},
  author={Stamenic, D and Rousseau, A and Essig, M and Gatault, P and Buchler, M and Filloux, M and Marquet, P and Pr{\'e}maud, A}
}

@article{belay2023joint,
  title={Joint modeling on serum creatinine and time to end stage of renal disease for chronic kidney disease patients under treatment at the University of Gondar Referral Hospital},
  author={Belay, Samson and Melese, Dessie and Muhammed, Kasim},
  journal={Health Science Reports},
  volume={6},
  number={9},
  pages={e1563},
  year={2023},
  publisher={Wiley Online Library}
}

\newpage
\section*{Appendix A}
\setcounter{equation}{0}
\renewcommand{\theequation}{A.\arabic{equation}}
\setcounter{table}{0}
\renewcommand{\thetable}{A. \arabic{table}}
\setcounter{figure}{0}
\renewcommand{\thefigure}{A. \arabic{figure}}

\begin{table}[h!]
\centering
\footnotesize
\caption{Classification of medications commonly used by study participants, grouped by therapeutic target and subdivided by mechanism of action. A binary covariate was created for each subgroup for modelling purposes.}\label{medications}
\begin{tabular}{llll}
\toprule
\textbf{Group} & \textbf{Subgroup} & \textbf{Medications} \\
\midrule

\multirow{8}{*}{\textbf{Autoimmune Conditions}}
  & \multirow{4}{*}{Corticosteroids}
    & Prednisolone \\
  & & Methylprednisolone sodium succinate \\
  & & Hydrocortisone \\
  & & Dexamethasone \\
\cmidrule{2-3}
  & Immunosuppressants & Mycophenolate mofetil \\
\cmidrule{2-3}
  & Immune Modulators & Hydroxychloroquine sulfate \\
\cmidrule{2-3}
  & B-Cell Therapy & Rituximab \\
\midrule

\multirow{4}{*}{\textbf{Hypertension \& Heart Disease}}
  & \multirow{2}{*}{Calcium Channel Blockers (CCB)}
    & Amlodipine \\
  & & Nifedipine \\
\cmidrule{2-3}
  & \multirow{2}{*}{ACE Inhibitors}
    & Lisinopril \\
  & & Enalapril \\
\bottomrule
\end{tabular}
\end{table}

\newcommand{\insig}{\textcolor{red}{\checkmark}}
\newcommand{\sig}{\checkmark}
\newcommand{\no}{$\times$}
\newcommand{\beststat}[1]{\textbf{\textcolor{red}{#1}}}

\begin{table}[H]
\centering
\footnotesize
\caption{Summary of joint model specifications and goodness-of-fit statistics.  
$\checkmark$: included and significant; $\checkmark^*$: included but non-significant;  
$\times$: excluded. WAIC: Watanabe–Akaike Information Criterion (lower is better); LPML: Log Pseudo-Marginal Likelihood (higher is better). Bold values indicate the best-performing model.}
\label{model_selection_criteria}
\begin{tabular}{lccccccc}
\toprule
& \textbf{M1} & \textbf{M2} & \textbf{M3} & \textbf{M4} & \textbf{M5} & \textbf{M6} & \textbf{M7} \\
\midrule
\multicolumn{8}{l}{{\textit{Longitudinal Outcome: log(Creatinine)}}} \\
Time & \checkmark & \checkmark & \checkmark & \checkmark & \checkmark & \checkmark & \checkmark \\
Age at entry (SAge) & \checkmark & \checkmark & \checkmark & \checkmark & \checkmark & \checkmark & \checkmark \\
Sex  & \checkmark & \checkmark & \checkmark & \checkmark & \checkmark & \checkmark & \checkmark \\
BMI $z$-score (BMIZ) & \checkmark & \checkmark & \checkmark & \checkmark$^*$ & \checkmark & \checkmark & \checkmark \\
\midrule
\multicolumn{8}{l}{{\textit{Survival Outcome: Hazard of composite kidney-related event}}} \\
Comorbidity (Comorb) & \checkmark$^*$ & \checkmark$^*$ & \checkmark$^*$ & \checkmark$^*$ & \checkmark$^*$ & \checkmark$^*$ & \checkmark$^*$ \\
Kidney condition (KidneyHist) & \checkmark$^*$ & \checkmark & \checkmark & \checkmark$^*$ & \checkmark & \checkmark & \checkmark \\
Corticosteroid (Cortico) & \checkmark & \checkmark & \checkmark & \checkmark & \checkmark & \checkmark & \checkmark \\
Immunosuppressant (Immuno) & \checkmark & \checkmark & \checkmark & \checkmark & \checkmark & \checkmark & \checkmark \\
Immune modulator (ImmMod) & \checkmark & \checkmark & \checkmark & \checkmark & \checkmark & \checkmark & \checkmark \\
B-cell therapy (BCell) & \checkmark$^*$ & \checkmark$^*$ & \checkmark$^*$ & \checkmark$^*$ & \checkmark$^*$ & \checkmark$^*$ & \checkmark$^*$ \\
Calcium channel blocker (CCB) & \checkmark & \checkmark & \checkmark & \checkmark & \checkmark & \checkmark & \checkmark \\
ACE inhibitor (ACEi) & \checkmark$^*$ & \checkmark$^*$ & \checkmark$^*$ & \checkmark$^*$ & \checkmark$^*$ & \checkmark$^*$ & \checkmark$^*$ \\
\midrule
\multicolumn{8}{l}{\textit{Association structure}} \\
Value (current log-creatinine, $\eta$) & \checkmark & \checkmark & \checkmark & \no & \no & \checkmark & \no \\
Slope (slope of $\eta$) & \checkmark$^*$ & \checkmark$^*$ & \no & \checkmark$^*$ & \no & \no & \checkmark$^*$ \\
Area (cumulative $\eta$) & \checkmark & \no & \no & \no & \checkmark & \checkmark & \checkmark \\
\midrule
\multicolumn{8}{l}{\textit{Model fit}} \\
WAIC & $-$1362.1 & $-$1354.0 & $-$1358.7 & $-$1259.5  & {\textbf{$-$1377.1}} & $-$1363.4 & $-$1362.5 \\
LPML & 195.7  & 171.6  & 174.1  & 116.9 & {\textbf{229.0}} & 139.7  & 188.2 \\
\bottomrule
\multicolumn{8}{l}{} \\
\end{tabular}
\end{table}

\end{document}